\documentclass[11pt,sort&compress]{elsarticle}
\makeatletter
\def\ps@pprintTitle{%
 \let\@oddhead\@empty
 \let\@evenhead\@empty
 \def\@oddfoot{\centerline{\thepage}}%
 \let\@evenfoot\@oddfoot}
\makeatother
\usepackage{amsmath}
\usepackage{accents}
\usepackage{amssymb}
\usepackage{mathrsfs}
\usepackage[LGR,T1]{fontenc}
\usepackage[latin1]{inputenc}
\usepackage[cal=boondox,calscaled=1]{mathalfa}

\DeclareMathAlphabet{\bbvar}{U}{BOONDOX-ds}{m}{n}
\DeclareMathAlphabet{\bbgreek}{U}{bbold}{m}{n}
\usepackage[defaultsans,osfigures,scale=0.9]{opensans}
\usepackage{graphicx}
\usepackage{psfrag}
\usepackage{pstool}
\usepackage{caption}
\usepackage{graphicx}%
\newcommand{\hook}{\text{\large{$\lrcorner$}}}
\usepackage{color}
\usepackage{hyperref}
\newcommand{\qq}[1]{``#1''} 

\newcommand{\di}{\mathrm{d}}
\usepackage{tensor}\newcommand{\ou}[3]{{#1}{}^{#2}{}_{#3}}
\newcommand{\uo}[3]{{#1}{}_{#2}{}^{#3}}

\newcommand{\ucheck}[1]{\underaccent{\check}{#1}}
\newcommand{\uepsilon}{\boldsymbol{\underaccent{\check}{\epsilon}}}
\newcommand{\oepsilon}{\boldsymbol{\hat{\epsilon}}}

\newcommand{\I}{\mathrm{i}} 
\newcommand{\E}{\mathrm{e}} 
\newcommand{\CC}{\mathrm{cc.}} 
\newcommand{\C}{\mathbb{C}}

\newcommand{\R}{\mathbb{R}}

\newcommand{\1}{\mathnormal{1}}
\newcommand{\0}{o}

\newenvironment{subalign}{\subequations\align}{\endalign\endsubequations}
\newcommand{\eref}[1]{(\ref{#1})}
\renewcommand{\b}{\bar}
\renewcommand{\bold}[1]{\boldsymbol{#1}}

\DeclareMathAlphabet{\sfit}{OT1}{fos}{sb}{it}
\DeclareMathAlphabet{\mathsf}{OT1}{fos}{sb}{n}

\begin{document}

\begin{abstract}
The covariant Hamiltonian formulation for general relativity is studied in terms of self-dual variables  on a manifold with an internal and lightlike boundary. At this inner boundary, new canonical variables appear: a spinor and a spinor-valued two-form that encode the entire intrinsic geometry of the null surface. At a two-dimensional cross-section of the boundary, quasi-local expressions for the generators of two-dimensional diffeomorphisms, time translations, and dilatations of the null normal are introduced and written in terms of the new boundary variables. In addition, a generalisation of the first-law of black-hole thermodynamics for arbitrary null surfaces is found, and the relevance of the framework for non-perturbative quantum gravity is stressed and explained. 

\end{abstract}
\title{New boundary variables for classical and quantum gravity on a null surface}
\author{Wolfgang Wieland}
\address{Perimeter Institute for Theoretical Physics\\31 Caroline Street North\\ Waterloo, ON N2L\,2Y5, Canada\\{\vspace{0.5em}\normalfont April 2017}
}
\date{Fall 2016}
\maketitle
{\tableofcontents}
\begin{center}{\noindent\rule{\linewidth}{0.4pt}}\end{center}
\section{Introduction: Why spinors?}

In covariant and non-perturbative quantum gravity \cite{Ashtekar, rovelli, thiemann, status,alexreview}, 
the quantum states of the gravitational field are built from gravitational Wilson lines for an $SU(2)$ (resp. $SL(2,\C)$) spin connection $A$, 
and the simplest and most elementary such excitations are given by the trace of the parallel transport around a loop $\alpha$,
\begin{equation}
\Psi_{\alpha}[A]=\mathrm{Tr}\bigg(\operatorname{Pexp}\Big(-\oint_\alpha A\Big)\bigg).
\end{equation}
Taking tensor products and superpositions of countably many such states, quantum geometries arise that represent distributional configurations of the gravitational field: In the naive semi-classical $\hbar\rightarrow 0$ limit, the semi-classical configuration of the inverse and densitised triad (the gravitational analogue of the Yang\,--\,Mills electric field) has support only along a one-dimensional fabric of loops \cite{contphas}. 

In the construction of these states, inner boundaries were neglected. If they are included, the Wilson lines can hit a two-dimensional boundary, where they create a surface charge. Now, the gravitational charge for internal frame rotations is spin (a spin connection  couples naturally to a spinor in the associated spin bundle), and the charges thus appearing are spinors that sit at two-dimensional boundaries. 
In the last couple of years, a new representation of quantum geometry emerged, where these pure-gravity spinors were taken as the fundamental configuration variables of the theory \cite{twist,komplexspinors,twistintegrals}. There is a classical phase space, Poisson brackets, constraints, and by now there are basically two proposals \cite{twistintegrals,Wieland:2014nka} and \cite{Wieland:2016aa} for how to formulate the dynamics of the quantum theory in terms of these spinors alone. An interpretation was missing, however, for what these spinors would be from the perspective of classical general relativity. This is an important problem, because it is a version of the most crucial open problem in the field, namely the question for how to regain the theory in the continuum (cf.\ \cite{Dittrich:2014ala}, which contains a very concise summary of the problem).
\vspace{1em}

From the viewpoint of classical general relativity, spinors 
play a fundamental role in the geometric analysis of null congruences and asymptotic null infinity \cite{penroserindler}. In non-perturbative quantum gravity, spinors appear as boundary variables at the open ends of gravitational Wilson lines. The purpose of this paper is to demonstrate that there is a more profound relation: 
We will study the Hamiltonian formulation of general relativity in terms of self-dual variables in a domain bounded by an inner null surface, and we will find, in fact, that the natural boundary variables for the Hamiltonian theory are given by a spinor $\ell^A$ and a spinor-valued two-form $\bold{\eta}_{Aab}$ intrinsic to the boundary. At a two-dimensional cross-section of the boundary, these spinors will contribute a corner term to the covariant pre-symplectic potential, and $\ell^A$ and $\bold{\eta}_{Aab}$ will be canonically conjugate variables (the boundary is three-dimensional, and in three dimensions a spinor is canonically conjugate to a spinor-valued surface density, hence a two-form). The resulting symplectic structure for these gravitational boundary variables, and the reality conditions that they have to satisfy turn out to be in one-to-one correspondence with the discrete structures that appear for the quantum-gravitational spinors at the open ends of gravitational Wilson lines \cite{komplexspinors}. 
 
In developing the field theoretical description of the boundary spinors, we will also develop further the covariant Hamiltonian formalism \cite{Wald:1999wa,WaldBHbook,Ashtekar:2008jw,Corichi:2015cqa} for complex Ashtekar variables in the presence of inner null boundaries, thereby generalising the isolated horizon framework \cite{isohorizon,Ashtekar:2001is,Ashtekar:aa,EngleNouiPerezPranzetti} to generic lightlike boundaries. We will study, in particular, the gauge symmetries of the system, which are $U(1)$ rotations of the boundary spinors, internal $SL(2,\C)$ frame rotations in the bulk and boundary, and four-dimensional diffeomorphisms that vanish at the corner. We will then also introduce the generators for dilatations of the null normal and two-dimensional diffeomorphisms at the corner, and they are not gauge transformations, but genuine Hamiltonian motions, whose {on-shell} generators define quasi-local expressions for energy and angular momentum. Finally, we will also introduce a quasi-local version of the first law, and explain the relevance of the framework for quantum gravity.

\vspace{1em}
The setup is the following: We study general relativity as a Hamiltonian system in a four-dimensional region $\mathcal{M}$, 
whose boundary has four components: an initial spatial hypersurface $\varSigma_0$, a final spatial hypersurface $\varSigma_1$, an inner lightlike cylinder $\mathcal{N}$ and a time-like outer cylinder $\mathcal{B}$. The three-surfaces $\varSigma_0$ and $\varSigma_1$ intersect this outer cylinder in outer corners $\mathcal{C}_0^{out}$ and $\mathcal{C}_1^{out}$ respectively. There are two further corners in the setup, which are the intersections of the inner null surface $\mathcal{N}$ with $\varSigma_0$ and $\varSigma_1$. We call them $\mathcal{C}_0^{in}$ and $\mathcal{C}_1^{in}$ respectively. The situation is summarised in \hyperref[fig1]{figure 1}.  All such corners have the topology of a two-sphere, and their orientation is chosen as follows: The three-dimensional boundary components $(\varSigma_0)^{-1}$, $\varSigma_1$, $\mathcal{B}$ and $\mathcal{N}$ inherit their orientation from the bulk, $(\mathcal{C}_1^{in})^{-1}$ and $\mathcal{C}_1^{out}$ inherit the orientation from $\varSigma_1$, and equally for the initial corners $(\mathcal{C}_0^{in})^{-1}$ and $\mathcal{C}_0^{out}$, which inherit their orientation from $\varSigma_0$, where e.g.\ $\mathcal{C}^{-1}$ denotes the manifold $\mathcal{C}$ equipped with opposite orientation, hence $\partial\mathcal{N}=(\mathcal{C}_0^{in})^{-1}\cup\mathcal{C}_1^{in}$. 

\vspace{1em}
Before we enter the construction, a few further remarks. In the following, we will use Penrose's index notation: Indices $a,b,c,\dots$ are abstract tensor indices labelling the sections of the tensor bundle over spacetime (or a submanifold in there). Sections of the spin bundle carry abstract and \emph{internal} spin-indices $A,B,C,\dots$ transforming under the fundamental spin $(\tfrac{1}{2},0)$ representation of $SL(2,\C)$. Primed indices $A',B',C',\dots$ transform under the complex conjugate spin $(0,\tfrac{1}{2})$ representation. The relation to \emph{internal} Minkowski indices $\alpha,\beta,\gamma,\dots$ is given by the intertwining and internal \emph{soldering forms}\footnote{An explicit matrix representation is given by the four-dimensional Pauli matrices $\ou{\sigma}{\rm{AA}'}{\mu}\equiv\sigma_\mu=(\bbvar{1},\sigma_1,\sigma_2,\sigma_3)$. Spinor indices are raised and lowered using the alternating epsilon spinors $\epsilon^{AB}$ and $\epsilon_{AB}$. Our conventions are the following: $\psi_A=\psi^B\epsilon_{BA}$ and $\psi^A=\epsilon^{AB}\psi_B$, equally for the complex conjugate representation.} $\ou{\sigma}{AA'}{\alpha}$, which provide the isomorphism between Lorentz vectors $V^\alpha$ and anti-hermitian\footnote{This is a consequence of our $(-$$+$$+$$+)$ sign  convention for the Minkowski metric $\eta_{\alpha\beta}$.} world-spinors $V^{AA'}$,
 \begin{align}
V^\alpha&=\frac{\I}{\sqrt{2}}\uo{\sigma}{AA'}{\alpha}V^{AA'},\\
V^{AA'}&=\frac{\I}{\sqrt{2}}\ou{\sigma}{AA'}{\alpha}V^{\alpha}.
\end{align}

\section{Action and variational principle for inner null boundaries}\label{sec2}
\subsection{Variables in the bulk}
Our fundamental gravitational configuration variables in the bulk are the tetrad $\ou{e}{\alpha}{a}$ and an $SO(1,3)$ connection $\ou{A}{\alpha}{\beta a}$. The vacuum field equations, which are Einstein's equations and the torsionless condition, follow from the Hilbert\,--\,Palatini action\begin{equation}
S_{\mathcal{M}}[A,e]=\frac{1}{16\pi G}\int_{\mathcal{M}}\ast\Sigma_{\alpha\beta}\wedge F^{\alpha\beta}.\label{bulkactn}
\end{equation}
$\Sigma_{\alpha\beta}$ is the Pleba\'{n}ski two-form
\begin{equation}
\Sigma_{\alpha\beta}=e_\alpha\wedge e_\beta,
\end{equation}
and $\ou{F}{\alpha}{\beta}=\di\ou{A}{\alpha}{\beta}+\ou{A}{\alpha}{\mu}\wedge\ou{A}{\mu}{\beta}$ is the field strength of the connection. The Hodge dual \qq{$\ast$} is taken in internal space: $\ast\Sigma_{\alpha\beta}=1/2\,\epsilon_{\alpha\beta\mu\nu}\Sigma^{\mu\nu}$.

In quantum gravity, the relevant gauge group is $SL(2,\C)$ rather than $SO(1,3)$. It is therefore useful to switch to $SL(2,\C)$ gauge covariant variables already at the level of the classical action,
\begin{equation}
S_{\mathcal{M}}[A,e]=\frac{\I}{8\pi G}\int_{\mathcal{M}}\Sigma_{AB}\wedge F^{AB}+\CC,\label{selfdualactn}
\end{equation}
where \qq{$\CC$} stands for the complex conjugate of all preceding terms (the action is real). Moreover,
\begin{equation}
\Sigma_{AB}=-\frac{1}{2}e_{AC'}\wedge\uo{e}{B}{C'}\label{Plebform}
\end{equation}
is the self-dual\footnote{The Hodge dual acts as $\ast\Sigma_{AB}=\I\Sigma_{AB}$.}  component 
of the Pleba\'nski two-form and 
\begin{equation}
\ou{F}{A}{B}=\di\ou{A}{A}{B}+\ou{A}{A}{C}\wedge\ou{A}{C}{B},
\end{equation}
is the curvature of the self-dual connection \cite{Sen:1982qb,newvariables}, which is the self-dual part of the $SO(1,3)$ connection  $\ou{A}{\alpha}{\beta a}$ one-form.

\vspace{0.5em}

Next, we have to add boundary terms and determine the boundary conditions for the outer and inner three-boundaries. The boundary term for the outer and timelike cylinder $\mathcal{B}$ is an extension of the Gibbons\,--\,Hawking\,--\,York boundary term for tetrad connection variables. This term has been first introduced by Obukhov \cite{obukhov}, and it can be written in the following form (see \cite{Bodendorfer:2013hla, komplex1} for references)
\begin{equation}
S_\mathcal{B}[A,e|z]=\frac{1}{8\pi G}\int_\mathcal{B}\ast\Sigma_{\alpha\beta}\wedge z^\alpha Dz^\beta,\label{GHY}
\end{equation}
where the internal, space-like and outwardly pointing four-vector $z^\alpha:z^\alpha z_\alpha=1$ defines the surface normal of the outer cylinder (the pull-back of the one-form  $z=z^\alpha e_\alpha$  to $\mathcal{B}$ vanishes). 
The boundary action \eref{GHY} also contains the vector-valued one-form $Dz^\alpha$, which is the gauge covariant exterior derivative
\begin{equation}
D z^\alpha=\di z^\alpha+\ou{A}{\alpha}{\beta}z^\beta
\end{equation}
on the boundary. The boundary conditions are that both the Pleba\'nski two-form and the internal normal are kept fixed in the variational problem, hence
\begin{equation}
(\varphi^\ast_{\mathcal{B}}\delta\Sigma_{\alpha\beta})_{ab}\equiv\delta\Sigma_{\alpha\beta\underleftarrow{ab}}=0,\quad\text{and}\quad\delta z^\alpha=0,
\end{equation}
where $\varphi^\ast_{\mathcal{B}}$ denotes the pull-back of the embedding $\varphi_\mathcal{B}:\mathcal{B}\hookrightarrow\mathcal{M}$ of the boundary into the bulk.

\begin{figure}[h]
\begin{center}
\psfrag{S0}{$\mathcal{C}_\1\;$}
\psfrag{S1}{$\mathcal{C}_\0\phantom{x}$}
\psfrag{N}{$\mathcal{N}\,$}
\psfrag{I}{$i_o$}
\psfrag{I+}{$\mathcal{I}^+$}
\psfrag{B}{$\mathcal{M}$}
\psfrag{R1}{$\varSigma_{\0}$}
\psfrag{R0}{$\varSigma_\1$}
\psfrag{Im}{$\mathcal{I}^+$}
\psfrag{i}{$\mathcal{B}\rightarrow i^\0$}
\psfrag{S}{$\varSigma_\1$}
\includegraphics[width=0.6\textwidth]{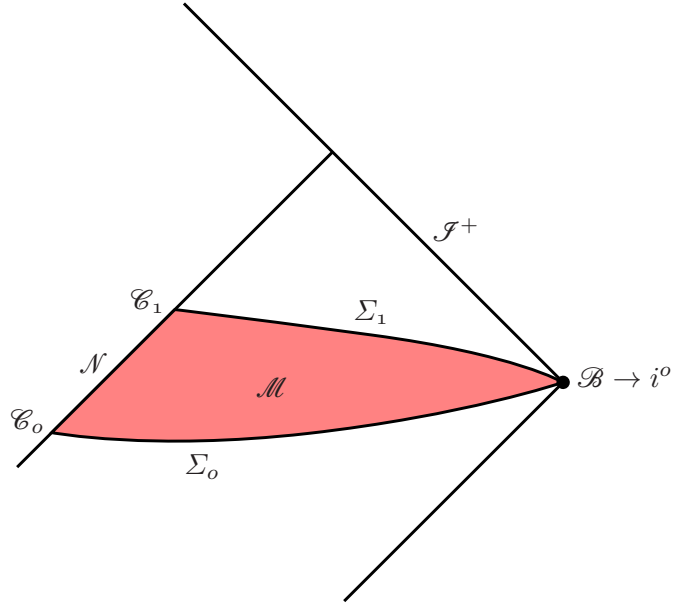}
\end{center}
\caption{We study the gravitational field in a subregion $\mathcal{M}$ as a Hamiltonian system on the covariant phase of the theory (for asymptotically flat spacetimes). The four-volume $\mathcal{M}$ is bounded by an inner and expanding null surface $\mathcal{N}$ reaching future null infinity, by partial Cauchy hypersurfaces $\varSigma_0$ and $\varSigma_1$, and by an outer time-like cylinder $\mathcal{B}$, which will be sent to spacelike infinity $i^o$.}\label{fig1}
\end{figure}

\subsection{Kinematical spin structures on a null surface}\label{sec2.1}
Consider then the inner null boundary, where we have to specify the variational principle and a suitable boundary term. The action \eref{bulkactn} treats the connection and the tetrad as independent variables, and this functional independence should be realised at the level of the boundary as well. This is an important hint for what the right boundary term should be, but there is a further helpful observation. Consider the Pleba\'nski two-form $\Sigma_{ABab}$ and the pull-back $\varphi_{\mathcal{N}}^\ast:T^\ast\mathcal{M}\hookrightarrow T^\ast\mathcal{N}$ to a three-surface $\mathcal{N}$. If the three-surface is null, the pull-back $(\varphi^\ast_{\mathcal{N}}\Sigma_{AB})_{ab}$ assumes a very simple algebraic form: it turns into the symmetrised tensor product of a spinor-valued two-form $\ou{\bold{\eta}}{A}{ab}\in\Omega^2(\mathcal{N}:\C^2)$ and a spinor $\ell^A\in\Omega^0(\mathcal{N}:\C^2)$, which are both intrinsic to $\mathcal{N}$. More explicitly,
\begin{equation}
(\varphi^\ast_\mathcal{N}\Sigma_{AB})_{ab}\equiv{\Sigma}_{AB\underleftarrow{ab}}=\ell_{(A}\bold{\eta}_{B)ab}.\label{glu}
\end{equation}
For any such spinors $({\bold{\eta}}_{Aab},\b\ell^{A'})$, we can define the following irreducible spin components, which all have an immediate geometric interpretation: The spin $(0,0)$ singlet
\begin{equation}
\boldsymbol{\varepsilon}_{ab}:=-\I\boldsymbol{\eta}_{Aab}\ell^A\label{areadef}
\end{equation}
defines the two-dimensional area two-form on the null hypersurface: If $\mathcal{C}$ is a two-dimensional submanifold $\mathcal{C}\subset\mathcal{N}$, the oriented and metrical area\footnote{The sign of $\boldsymbol{\varepsilon}$ may be used to distinguish infalling from outgoing null shells.} of $\mathcal{C}$ will be given by the integral
\begin{equation}
\operatorname{Ar}[\mathcal{C}]=\int_{\mathcal{C}}\boldsymbol{\varepsilon}.
\end{equation}
For the area to be real, we have to satisfy the \emph{reality condition}
\begin{equation}
C=\bold{\eta}_{Aab}\ell^A+\CC=0.\label{realcond}
\end{equation}
Next, we can also build the spin $(\tfrac{1}{2},\tfrac{1}{2})$ vector component
\begin{equation}
\ell^a=\I\uo{e}{AA'}{a}\ell^A\bar{\ell}^{A'},\label{elldef}
\end{equation}
which defines the unique null direction
\begin{equation}
\ell^a\in T\mathcal{N}:\ell_a\ell^a=0
\end{equation}
of the null hypersurface $\mathcal{N}$.
In fact, the entire \emph{intrinsic} geometry of the null surface can be reconstructed from the \emph{boundary spinors}
\begin{equation}
{Z}^{\mathfrak{a}}=\begin{pmatrix}\ell^A\\\bold{\bar{\eta}}_{A'ab}\end{pmatrix},\qquad\bar{Z}_{\mathfrak{a}}:=(\bold{\eta}_{Aab},\bar{\ell}^A)\label{bndryspin}
\end{equation}
alone. The proof can be found in \cite{Wieland:2016aa}, and I repeat it in \hyperref[appdxA]{appendix A}. The result is the following: There always exists a normalised dyad $(k_A,\ell_A):k_A\ell^A=1$ in $\C^2$ and an associate triad $(k_a,m_a,\bar{m}_a)\in T^\ast_\C\mathcal{N}$, where the boundary-intrinsic one-form $k_a$ is real and $m_a$ is complex, such that we obtain  the decomposition
\begin{equation}
\bold{\eta}_{Aab}=2\ell_Ak_{[a}\bar{m}_{b]}+2k_A\bar{m}_{[a}m_{b]},\label{mudecomp}
\end{equation}
for $k_a\ell^a=-1$ and $m_a\ell^a=0$ (see \hyperref[appdxA]{appendix A} for the details).  The dyad $(m_a,\bar{m}_a)$ plays an important role. It defines a degenerate signature $(0$$+$$+)$ metric $q_{ab}$ on $\mathcal{N}$, which is simply given by
\begin{equation}
q_{ab}=2m_{(a}\bar{m}_{b)}.
\end{equation}
If the glueing conditions \eref{glu} are satisfied (see again \hyperref[appdxA]{appendix A} for the details), this three-metric $q_{ab}\in T^\ast\mathcal{N}\otimes T^\ast\mathcal{N}$ is nothing but the pull-back of the four-dimensional  spacetime metric $g_{ab}=\eta_{\alpha\beta}\ou{e}{\alpha}{a}\ou{e}{\beta}{b}$ to the boundary
\begin{equation}
q_{ab}=g_{\underleftarrow{ab}}.
\end{equation}
Finally, from any such boundary spinor $Z^{\mathfrak{a}}$, one can also reconstruct the pull-back of the entire four-dimensional soldering form to the boundary, namely
\begin{equation}
(\varphi^\ast_{\mathcal{N}} e^\alpha)_a\equiv\ou{e}{AA'}{\underleftarrow{a}}=
-\I \ell^A\bar{\ell}^{A'}k_a+\I\ell^A\bar{k}^{A'}\bar{m}_a+\I k^A\bar{\ell}^{A'}m_a.\label{pullbcktetra}
\end{equation}

Notice, however, that there is no canonical such decomposition. The gauge transformations
\begin{equation}\left.\begin{split}
\ell^A&\longrightarrow\E^{+\frac{\lambda+\I\phi}{2}}\ell^A,\\
k^A&\longrightarrow\E^{-\frac{\lambda+\I\phi}{2}}(k^A+\bar\zeta\ell^A),
\end{split}\,
\begin{split}
\phantom{\E^{\frac{\lambda}{2}}}k_a&\longrightarrow \E^{-\lambda}(k_a+\bar\zeta m_a+\zeta\bar{m}_a),\\
\phantom{\E^{\frac{\lambda}{2}}}m_a&\longrightarrow\E^{\I\phi}m_a,
\end{split}\;\right\}\label{gaugespin}
\end{equation}
clearly leave both the Pleba\'nski two-form \eref{mudecomp} and the tetrad \eref{pullbcktetra} invariant (for \emph{local} gauge parameters $\zeta\in\C$ and $\lambda$, $\varphi\in\R$). At the level of the boundary spinor $Z^{\mathfrak{a}}$, the gauge symmetries \eref{gaugespin} simplify: Only the dilatations and $U(1)$ rotations survive:
\begin{equation}
(\bold{\eta}_{Aab},\b\ell^{A'})\longrightarrow(\E^{-\frac{\lambda+\I\phi}{2}}\bold{\bar\eta}_{Aab},\E^{+\frac{\lambda-\I\phi}{2}}\b\ell^{A'}).\label{complxgauge}
\end{equation}

Before proceeding further, we summarise. {The main message so far is that the primary variables at the null surface are the boundary spinors $\ell^A\in\Omega^0(\mathcal{N}:\C^2)$ and $\ou{\bold{\eta}}{A}{ab}\in\Omega^2(\mathcal{N}:\C^2)$ modulo gauge \eref{complxgauge}.  The area two-form $\bold{\varepsilon}_{ab}$, the equivalence class $[\ell^a]$ of null generators and the signature $(0$$+$$+)$ metric $q_{ab}$ are secondary, for they can be all reconstructed from $\ell^A$ and $\bold{\eta}_{Aab}$ alone.}

\subsection{Spin connection coefficients, extrinsic curvature}\label{sec2.2}
In the last section, we saw that the boundary spinor $\bar{Z}_{\mathfrak{a}}=(\bold{\eta}_{Aab},\bar{\ell}^{A'})$ describes the entire \emph{intrinsic geometry} of the null surface. The next step concerns a notion of \emph{extrinsic curvature}, as measured by an equivalence class of spin connections on the boundary. If we view the boundary as embedded into the bulk, the situation is clear: Given the entire four-dimensional geometry, there is the preferred Levi-Civita spin connection and its exterior covariant derivative $\nabla_a$, which maps spinor valued $p$-forms $\phi^{AB\dots A'B'\dots}$ into $(p+1)$-forms $\nabla\phi^{AB\dots A'B'\dots}$. The derivative annihilates the internal $\epsilon$-spinors (e.g.\ $\nabla_a\epsilon_{AB}=0$), and it is completely determined by the torsionless condition
\begin{equation}
\nabla\Sigma^{AB}=0\Leftrightarrow \nabla_{[a}\ou{\Sigma}{AB}{cd]}=0.\label{trsnless}
\end{equation}
This derivative $\nabla_a$ can be naturally pulled-back to the boundary, obtaining the exterior covariant derivative
\begin{equation}
D(\varphi^\ast_{\mathcal{N}}{\phi}^{AB\dots A'B'\dots}):=\varphi^\ast_{\mathcal{N}} \nabla{\phi}^{AB\dots A'B'\dots},\label{Ddef}
\end{equation}
which maps spinor valued $p$-forms $\psi^{AB\dots A'B'\dots}$ on $\mathcal{N}$ into $(p+1)$-forms on $\mathcal{N}$. The idea is then to take $D_a$ as the prime object, and ask how much it knows about the extrinsic curvature. In fact, the intrinsic geometry of the null surface does not fix $D_a$ completely, the missing data being the extrinsic curvature. This is mirrored by the situation at a spacelike hypersurface, where the Ashtekar connection \cite{newvariables} depends on both the intrinsic and extrinsic geometry.\vspace{1em}

 Now, the intrinsic geometry of the null surface is determined completely by the boundary spinor $Z^{\mathfrak{a}}$ (as in \eref{bndryspin} above), and our strategy is to compute its exterior covariant derivative with respect to $D_a$, and identify those components of $DZ^{\mathfrak{a}}$ that determine the extrinsic curvature provided the reality conditions \eref{realcond} and the pull-back of the torsionless condition \eref{trsnless} are satisfied. Consider then an adapted spin basis $(k_A,\ell_A):k_A\ell^A=1$ (as in equation \eref{mudecomp} above), and decompose the covariant derivative into complex spin coefficients $\gamma_a$, $\psi_a$ and $\varphi_a$, such that
\begin{subalign}
D_a\ell^A&=+\gamma_a\ell^A+\psi_ak^A,\\
D_ak^A&=-\gamma_ak^A-\phi_a\ell^A.
\end{subalign}
We can now go back to the glueing condition \eref{glu} and take the pull-back of the torsionless condition \eref{trsnless} to the boundary. A straightforward calculation yields
\begin{align}\nonumber
\varphi^\ast_{\mathcal{N}}(\nabla\Sigma_{AB})=D(\bold{\eta}_{(A}\ell_{B)})&=
\left(\di\bold{\mu}+2\bold{\mu}\wedge \gamma-\I\,\bold{\varepsilon}\wedge\psi\right)\ell_A\ell_B+\\
&\quad+\left(\I\,\di\bold{\varepsilon}+2\bold{\mu}\wedge\psi\right)\ell_{(A}k_{B)}+\I\,\bold{\varepsilon}\wedge\psi\,k_Ak_B,\label{trsnless2}
\end{align}
where
\begin{equation}
\bold{\mu}_{ab}=2k_{[a}\bar{m}_{b]},\quad\bold{\varepsilon}_{ab}=2\I m_{[a}\bar{m}_{b]}
\end{equation}
are the two-forms as in equation \eref{mudecomp} above. The vanishing of \eref{trsnless2} is therefore  equivalent to
\begin{subequations}
\begin{align}
\di\bold{\mu}+2\bold{\mu}\wedge\gamma-\I\bold{\varepsilon}\wedge\phi=0,\\
\vartheta_{(\ell)}k\wedge\bold{\varepsilon}+2\I\bold{\mu}\wedge\psi=0,\label{expnstrsn}\\
\bold{\varepsilon}\wedge\psi=0,\label{geodesty}
\end{align}\label{trsnless3}%
\end{subequations}%
where we have introduced the expansion $\vartheta_{(\ell)}$, which is defined intrinsically on $\mathcal{N}$ by the exterior derivative of the area two-form
\begin{equation}
\di\bold{\varepsilon}=-\vartheta_{(\ell)}k\wedge\bold{\varepsilon}\in \Omega^3(\mathcal{N}:\R).
\end{equation}

Before we proceed, a few consistency checks to get an intuition for these equations: Equation \eref{expnstrsn} determines the expansion $\vartheta_{(\ell)}$ as a function $\vartheta_{(\ell)}=-2\ell_A\bar{m}^aD_a\ell^A$ of the spin coefficients, while \eref{geodesty} is the same as to say $\ell^a\psi_a=0$, which implies, in turn, $\ell^aD_a\ell^A\sim\ell^A$. The internal null vector $\ell^\alpha\equiv\I\ell^A\bar{\ell}^A$ is therefore parallel along the null surface, which is well expected, since the integral curves of the null generators of a null surface are always geodesics.
\vspace{1em}

Our next goal is to translate the pull-back of the torsionless condition \eref{trsnless2} into constraints on the boundary spinors and their derivatives. First of all, we look at the exterior covariant derivative of $\bold{\eta}_{Aab}$. Going back to the torsionless condition \eref{trsnless3} and the decomposition \eref{mudecomp} of $\bold{\eta}_{Aab}$ in terms of the spin dyad $(k_A,\ell_A):k_A\ell^A=1$, we get
\begin{align}\nonumber
D\bold{\eta}_A&=\di\bold{\mu}_A+\I\,\di\bold{\varepsilon}\,k_A+\bold{\mu}\wedge\gamma\,\ell_A+\bold{\mu}\wedge\psi\, k_A+\\
\nonumber &\qquad-\I\,\bold{\varepsilon}\wedge\gamma\,k_A-\I\,\bold{\varepsilon}\wedge\phi\,\ell_A=\\
&=-\bold{\eta}_A\wedge\gamma-\frac{\I}{2}\bold{\varepsilon}\wedge k\,\vartheta_{(\ell)}k_A
=-\bold{\eta}_A\wedge\Big(\gamma+\frac{1}{2}\vartheta_{(\ell)} k\Big).\label{extderiv1}
\end{align}
The last line suggests to introduce the following complex-valued one-form on the boundary
\begin{equation}
\omega_a=\gamma_a+\frac{1}{2}\vartheta_{(\ell)} k_a\in T^\ast_\C\mathcal{N}.
\end{equation}
The exterior covariant derivative of $\ell^A$ can be then written in terms of $\omega_a$ and a shift $\ou{\psi}{A}{a}$, namely
\begin{equation}
D_a\ell^A=\omega_a\ell^A+\ou{\psi}{A}{a},\label{extderiv2}
\end{equation}
where we introduced the spinor-valued one-form
\begin{equation}
\ou{\psi}{A}{a}=\psi_a k^A-\frac{1}{2}\vartheta_{(\ell)}k_a\ell^A.
\end{equation}
We can then collect both exterior covariant derivatives of the boundary spinors, namely \eref{extderiv1} and \eref{extderiv2}, and write them together as
\begin{subequations}\begin{align}
(D_a-\omega_a)&\wedge\ell^A=\ou{\psi}{A}{a},\label{spineom1}\\
(D+\omega)&\wedge\bold{\eta}_A=0.\label{spineom2}
\end{align}\label{spineoms}%
\end{subequations}%
The reality conditions \eref{realcond} and the pull-back of the torsionless condition \eref{trsnless2} translate now into simple algebraic constraints on $\bold{\eta}_{Aab}$ and $\ou{\psi}{A}{a}$. First of all, there is the reality condition $C=0$, which is conserved along the null surface only if $\di C=0$, which is the same as to say
\begin{equation}
\di C=0\Rightarrow C':=\bold{\eta}_{A}\wedge\psi^A+\CC=0.\label{realcond2}
\end{equation}
Next, there is the torsionless equation \eref{trsnless}, which turns into
\begin{equation}
C_{AB}=\bold{\eta}_{(A}\wedge\psi_{B)}=0.\label{threetrsn}
\end{equation}

The one-forms $\omega_a$ and $\ou{\psi}{A}{a}$ play an important role on the covariant phase space of the theory: We will demonstrate in the next section that $\omega_a$ and $\ou{\psi}{A}{a}$ can be seen as the gravitational configuration variables on a null surface, while their conjugate moments are given by the area two-form $\bold{\varepsilon}_{ab}$ and the spinor-valued two-form $\bold{\eta}_{Aab}$. In addition to their importance for the Hamiltonian framework, $\omega_a$ and $\ou{\psi}{A}{a}$ have a clear geometric interpretation as well, for they determine a notion of extrinsic curvature on the null surface. This can be seen as follows: Consider first the pull-back of the torsionless equation $\ou{T}{\alpha}{ab}=2\nabla_{[a}\ou{e}{\alpha}{b]}=0$ (for a generic $SO(1,3)$ connection $\ou{A}{\alpha}{\beta a}$) to the null surface, namely
\begin{equation}
(\varphi^\ast_{\mathcal{N}}T^\alpha)_{ab} = 2 D_{[a}(\varphi^\ast_{\mathcal{N}} e^\alpha)_{b]}\stackrel{!}{=}0.\label{trsnbndry}
\end{equation}
It is easy to see that given the pull-back of the tetrad $\varphi^\ast_{\mathcal{N}}e^\alpha$ on the boundary, this equation has no unique solution for $D_a$ on $\mathcal{N}$. The difference between any two such covariant derivatives $D_a$ and $D^\circ_a$ defines an $\mathfrak{so}(1,3)$-valued one-form $\ou{K}{\alpha}{\beta a}$, which is defined for any internal four-vector $V^\alpha$ by
\begin{equation}
({D}_a-D^\circ_a)V^\alpha=\ou{K}{\alpha}{\beta a}V^\beta.
\end{equation}
On a spacelike hypersurface $\varSigma$, there is a natural reference connection (a natural origin) in the affine space of $SL(2,\C)$ connections on $\varSigma$, which is given by the intrinsic $SU(2)$ Levi-Civita connection. The difference tensor $\ou{K}{\alpha}{\beta a}$ is then canonical, and it is determined completely by the extrinsic curvature $K_{ab}$ of the spacelike hypersurface. On a null surface, the situation is more subtle: The space of connections has no preferred such origin \cite{AshtekarNullInfinity}, and the closest thing that determines the extrinsic geometry of the null surface is not the difference tensor $\ou{K}{\alpha}{\beta a}$, but rather the equivalence class $[D_a]$ of all $SO(1,3)$ connections on $\mathcal{N}$ under the equivalence relation
\begin{equation}
D_a\sim \tilde{D}_a\Leftrightarrow\left\{
\begin{split}
&D_a\ell^\alpha=\tilde{D}_a\ell^\alpha,\;\text{and}\\
&D_{[a}(\varphi^\ast_{\mathcal{N}} e^\alpha)_{b]}=\tilde{D}_{[a}(\varphi^\ast_{\mathcal{N}} e^\alpha)_{b]}=0.
\end{split}\right.\label{eqidef}
\end{equation}
A straightforward calculation shows that the difference $D_a-\tilde{D}_a=[K_a,\cdot]$ for any two such elements $D_a\sim\tilde{D}_a$ must be of the following general form
\begin{equation}
\ou{K}{\alpha\beta}{a}=2\ell^{[\alpha}m^{\beta]}\left(f\bar{m}_a+\chi m_a\right)+\CC,
\end{equation}
where $f:\mathcal{N}\rightarrow\R$ is real and $\chi$ is complex, while $(\ell^\alpha,m^\alpha,\bar{m}^\alpha)$ denotes an \emph{internal} Newman\,--\,Penrose triad, with respect to which the pull back of the tetrad assumes the same form
\begin{equation}
\ou{e}{\alpha}{\underleftarrow{a}}=(\varphi^\ast_{\mathcal{N}}e^\alpha)_a=-\ell^\alpha k_a+\bar{m}^\alpha m_a+m^\alpha\bar{m}_a
\end{equation} as in \eref{pullbcktetra} above.
In terms of the spin connection, we split $\ou{K}{\alpha}{\beta a}\equiv\ou{K}{AA'}{BB' a}$ into its self-dual and anti-self-dual components $\ou{K}{AA'}{BB' a}=\delta^{A'}_{B'}\ou{K}{A}{Ba}+\CC$, the self-dual part being given explicitly by
\begin{equation}
\ou{K}{A}{Ba}=-\ell^A\ell_B\left(f\bar{m}_a+\chi m_a\right).\label{difftens}
\end{equation}
It is now quite immediate to see that the potentials $\omega_a$ and $\ou{\psi}{A}{a}$ as in \eref{spineoms}, determine the connection $D_a$ only up to such a difference tensor: Equation \eref{spineom1} determines the connection up to a term $\ell^A\ell_B(f_ok_a+f\bar{m}_a+\chi m_a)$. From equation \eref{spineom2} it follows then that $f_o=0$ and the torsionless condition restricts $f$ to be real: $f=\bar{f}$, hence the difference tensor assumes again the generic form of equation \eref{difftens}. In other words, the equivalence class $[D_a]$ of $SO(1,3)$ connections on $\mathcal{N}$ can be characterised uniquely by a pair $[\omega_a,\ou{\psi}{A}{a}]$, and we write, therefore
\begin{equation}
[D_a]=[\omega_a,\ou{\psi}{A}{a}].
\end{equation}
Notice, however, that there is still a residual gauge freedom: We can always shift $\omega_a$ in \eref{spineoms} by a term $c\,\bar{m}_a$ (for some $c:\mathcal{N}\rightarrow\C)$ without changing \eref{spineom2}. This shift in $\omega_a$ can be then made undone by a corresponding shift of $\ou{\psi}{A}{a}$ without changing the covariant derivative $D_a\ell^A$ in \eref{spineom1}. The residual gauge freedom is, therefore,
\begin{equation}
\omega_a\sim\omega_a+c\,\bar{m}_a,\quad\ou{\psi}{A}{a}\sim\ou{\psi}{A}{a}-c\,\ell^A\bar{m}_a.
\end{equation}

Before we proceed to the next section, let me  summarise. {The main message of this section is that the exterior covariant derivatives \eref{spineoms} of the boundary spinors $\ell^A$ and $\bold{\eta}_{Aab}$ are completely characterised by $\omega_a$, which is a complex-valued one-form, and $\ou{\psi}{A}{a}$, which is a spinor-valued one-form, both intrinsic to $\mathcal{N}$. In addition, we also saw that $\omega_a$ and $\ou{\psi}{A}{a}$ determine an equivalence class $[D_a]=[\omega_a,\ou{\psi}{A}{a}]$ of torsionless spin connections on the boundary, and any such equivalence class measures the extrinsic curvature of the null surface, which is one of the canonical coordinates on the covariant phase space (more about this below). We will see, in fact, that $[\omega_a,\ou{\psi}{A}{a}]$ has conjugate moments given by a pair $(\bold{\varepsilon}_{ab},\bold{\eta}_{Aab})$ consisting of the area two-form $\bold{\varepsilon}_{ab}$ \eref{areadef}  and the spinor-valued two-form $\bold{\eta}_{Aab}$ \eref{glu}.}

\subsection{Boundary term and boundary equations of motion}
The goal of this section is to determine the boundary term and the boundary conditions at the inner null surface. Such a boundary term is needed, because otherwise the variational problem is ill-posed: The variation of the action yields the Einstein equations
\begin{align}
\ou{F}{A}{B}\wedge e^{BA'}=0,\qquad
\nabla\Sigma_{AB}=0\label{Eeq}
\end{align}
in the bulk, while at the inner three-boundary, we are left with the remainder
\begin{equation}
\frac{\I}{8\pi G}\int_{\mathcal{N}}\Sigma_{AB}\wedge\delta A^{AB}+\CC\label{remndr}
\end{equation}
The boundary condition is that the inner three-boundary $\mathcal{N}$ is null. This is a condition on the intrinsic geometry of $\mathcal{N}$. The extrinsic geometry, which is encoded in the connection, is left free to fluctuate around solutions that satisfy the boundary condition. We thus need a boundary term to cancel the $\delta A^{AB}$ connection variation from the bulk. The remainder \eref{remndr} is linear in the connection, as is the exterior derivative $D_a$, which acts on spinor-valued $p$-forms, such as $\ell^A$ and $\bold{\eta}_{Aab}$. The exterior covariant derivative maps $p$-forms into $(p+1)$-forms, the integrand must be a three-form, hence the natural boundary kinetic term is given by the integral
\begin{equation}
\frac{\I}{8\pi G}\int_{\mathcal{N}}\bold{\eta}_A\wedge D\ell^A+\CC
\end{equation}
The coupling constant in front has been chosen already such that the entire bulk plus boundary action is stationary provided both the Einstein equations \eref{Eeq} and the \emph{glueing conditions}
\begin{equation}
(\varphi^\ast_{\mathcal{N}}\Sigma^{AB})_{ab}=\ou{\Sigma}{AB}{\underleftarrow{ab}}=\ell_{(A}\bold{\eta}_{B)ab}\label{glucond}
\end{equation}
are satisfied. As we have seen in \hyperref[sec2.1]{section 2.1.\ }above, imposing the glueing conditions \eref{glucond} together with the reality conditions \eref{realcond} imposes that the boundary is null. Finally, we may also add the potentials $\omega_a$ and $\ou{\psi}{A}{a}$ to the action such that the variation with respect to the boundary spinor $\bar{Z}_{\mathfrak{a}}=(\bold{\eta}_{Aab},\bar\ell^{A'})$ returns the boundary equations of motion \eref{spineoms}. The resulting boundary action is therefore given by
\begin{equation}
S_{\mathcal{N}}[A|\bold{\eta},\ell|\omega,\psi]=\frac{\I}{8\pi G}\int_{\mathcal{N}}\big(\bold{\eta}_A\wedge \left(D-\omega\right)\ell^A-\bold{\eta}_A\wedge \psi^A\big)+\CC
\end{equation}
Notice also that the boundary spinor $({\bold\eta}_{Aab},\b\ell^{A'})$ is defined only modulo dilatations and $U(1)$ gauge transformations \eref{complxgauge}. Upon adding $(\omega_a,\ou{\psi}{A}{a})$ to the action, we have, in fact, turned these transformations into actual symmetries of the boundary action, the gauge transformations being
\begin{subequations}\begin{align}
({\bold\eta}_{Aab},\b\ell^{A'})&\longrightarrow(\E^{-\frac{\lambda+\I\phi}{2}}\bold{\eta}_{Aab},\E^{+\frac{\lambda-\I\phi}{2}}\b\ell^{A'}),\label{complxtrafo1}\\
(\omega_a,\ou{\psi}{A}{a})&\longrightarrow(\omega_a+\tfrac{1}{2}\partial_a(\lambda+\I\phi),\E^{+\frac{\lambda+\I\phi}{2}}\ou{\psi}{A}{a})\label{complxtrafo2},
\end{align}\label{complxtrafo}\end{subequations}
such that $\omega_a$ can be interpreted as a complexified $U(1)$ gauge connection.

The entire action is then the sum of contributions from the bulk and boundary, namely
\begin{align}\nonumber
S[A,e|\bold{\eta},\ell|\omega,\psi]=&\frac{\I}{8\pi G}\int_{\mathcal{M}}\Sigma_{AB}\wedge F^{AB}+\\
&\quad+\frac{\I}{8\pi G}\int_{\mathcal{N}}\left[\bold{\eta}_A\wedge \left(D-\omega\right)\ell^A-\bold{\eta}_A\wedge \psi^A\right]+\CC\label{fullactn}
\end{align}
If we take into account also the outer cylinder approaching spacelike infinity (see \hyperref[fig1]{figure 1} for an illustration) the Gibbons\,--\,Hawking\,--\,York boundary term \eref{GHY} for the outer cylinder should be included to the action as well. 

Finally, we can now also specify the boundary conditions at the inner null surface: The boundary potentials  $\omega_a$ and $\ou{\psi}{A}{a}$ are held fixed, the boundary spinor $({\bold\eta}_{Aab},\b\ell^{A'})$ is free to fluctuate on $\mathcal{N}$, while the variation of $\ell^A$ vanishes at the inner corners $\partial\mathcal{N}=(\mathcal{C}_0^{in})^{-1}\cup\mathcal{C}_1^{in}$. In summary,
\begin{equation}
\text{on $\mathcal{N}$}:\delta\omega_a=0=\delta\ou{\psi}{A}{a},\, 
\text{on $\partial\mathcal{N}$}:\delta\ell^A=0.
\end{equation}
The equations of motion follow then from the variation of the action provided the boundary conditions are satisfied. We get the Einstein equation \eref{Eeq} in the bulk, and additional equations of motion along the null boundary: The glueing conditions \eref{glucond} linking the bulk with the boundary and, in addition, the boundary equations of motion \eref{spineoms}, which determine the exterior covariant derivatives of the boundary spinor $\bar{Z}_a=(\bold{\eta}_{Aab},\b\ell^{A'})$.


\section{Null surfaces as Hamiltonian subsystems}\label{sec3}
\subsection{Pre-symplectic potential}\label{sec3.1}
Having specified the boundary term, we can now determine the covariant symplectic potential $\Theta_{\partial\mathcal{M}}$, which is obtained from the first variation of the action,
\begin{equation}
\delta S=\text{EOM}\cdot\delta+\Theta_{\partial\mathcal{M}}(\delta).
\end{equation}
Notice that there are equations of motion both in the bulk and boundary, namely the Einstein equations \eref{Eeq} in the bulk and the boundary equations of motion \eref{spineoms} for the boundary spinor $\bar{Z}_{\mathfrak{a}}=(\bold{\eta}_{Aab},\b\ell^{A'})$. The boundary action contains the covariant differential $D\ell^A$, its variation yields a boundary term at \emph{the boundary of the boundary}, which are the inner corners $\partial\mathcal{N}=(\mathcal{C}_0^{in})^{-1}\cup\mathcal{C}_1^{in}$. On shell, the first variation of the action \eref{fullactn} contains, therefore, contributions from both the three boundary and the two-dimensional corners, namely\footnote{For simplicity, we assume here that the variation vanishes at the outer corner. The more general case will be studied below.}
\begin{align}
\delta S\approx&+\left[\frac{\I}{8\pi G}\int_{\mathcal{C}_1^{in}}(\bold{\eta}_A\delta\ell^A-\CC)+\frac{\I}{8\pi G}\int_{\varSigma_1}(\Sigma_{AB}\wedge\delta A^{AB}-\CC)\right]+\nonumber\\
&-\left[\frac{\I}{8\pi G}\int_{\mathcal{C}_0^{in}}(\bold{\eta}_A\delta\ell^A-\CC)+\frac{\I}{8\pi G}\int_{\varSigma_0}(\Sigma_{AB}\wedge\delta A^{AB}-\CC)\right]+\nonumber\\
&-\frac{\I}{8\pi G}\int_{\mathcal{N}}(\bold{\eta}_A\ell^A\wedge\delta\omega-\bold{\eta}_A\wedge \delta\psi^A-\CC),
\end{align}
where $\approx$ means equality up to terms that are constrained to vanish provided the equations of motion in bulk \eref{Eeq} and boundary \eref{spineoms} are satisfied. 

The boundary splits into four pieces, and so does the covariant pre-symplectic potential. The requirement that the pre-symplectic potentials be all invariant under local $SL(2,\C)$ frame rotations (this point will be discussed more carefully below) implies that the contributions from the inner and outer corners should be added to the pre-symplectic potential on\footnote{For definiteness, $\varSigma=\varSigma_1$, the situation for the initial and partial Cauchy hypersurface is completely analogous.}  $\varSigma$. With \qq{$\bbvar{d}$} denoting the exterior functional differential, we obtain, therefore
\begin{align}\nonumber
\Theta_{\varSigma}=&\frac{\I}{8\pi G}\int_\varSigma\left(\Sigma_{AB}\wedge\bbvar{d}A^{AB}-\CC\right)+\\
&+\frac{\I}{8\pi G}\int_{\mathcal{C}_{\text{\it in}}}\left(\bold{\eta}_A\bbvar{d}\ell^A-\CC\right)-\frac{1}{8\pi G}\int_{\mathcal{C}_{\text{\it out}}}\ast\Sigma_{\alpha\beta}z^\alpha\bbvar{d}z^\beta,\label{actnpot}
\end{align}
where we have also taken into account the additional contribution from the first variation of the Gibbons\,--\,Hawking\,--\,York boundary term \eref{GHY}. The contribution from the inner null surface, on the other hand, is given by
\begin{equation}
\Theta_{\mathcal{N}}=-\frac{\I}{8\pi G}\int_{\mathcal{N}}\big(\bold{\eta}_A\ell^A\wedge\bbvar{d}\omega+\bold{\eta}_A\wedge\bbvar{d}\psi^A-\CC\big).\label{nullpot}
\end{equation}
The corresponding Poisson brackets on $\varSigma$ (at the pre-symplectic or kinematical level) are those for the original self-dual Ashtekar variables \cite{Ashtekar,newvariables} plus additional Poisson brackets at the inner corner, the only non-trivial\footnote{There are also the complex conjugate Poisson brackets for the primed variables $\ou{\bar{\Sigma}}{A'}{B'ab}$, $\ou{\bar{A}}{A'}{B'a}, \bold{\bar\eta}_{A'ab}, {\bar{\ell}}^{A'}$, which are obtained from \eref{Poiss1} and \eref{Poiss2} by complex conjugation. The primed and unprimed sectors commute under the Poisson bracket.} Poisson brackets being
\begin{subequations}\begin{align}
\big\{\Sigma_{Abab}(x),\ou{A}{CD}{c}(x')\big\}_{\varSigma}&=-8\pi\I\,G\,\uepsilon_{abc}\,\delta^{(C}_{A}\delta^{D)}_B\,\delta^{(3)}_\varSigma(x,x'),\\
\big\{\bold{\eta}_{Aab}(z),\ell^B(z')\big\}_{\mathcal{C}}&=-8\pi\I\,G\,\uepsilon_{ab}\,\delta^B_A\,\delta^{(2)}_{\mathcal{C}}(z,z')\label{CornerBrckt},
\end{align}\label{Poiss1}\end{subequations}
where $\uepsilon_{abc}$ (resp. $\uepsilon_{ab}$) is the canonical and \emph{metric independent} inverse Levi-Civita density on the partial Cauchy surface $\varSigma$ (resp. the inner corner $\mathcal{C}$), and $\delta_{\varSigma}^{(3)}(\cdot,\cdot)$ (resp. $\delta_{\mathcal{C}}^{(2)}(\cdot,\cdot)$) is the Dirac distribution (a density of weight one) on $\varSigma$ (resp. $\mathcal{C}$).

At the inner null surface, we can deduce Poisson commutation relations as well, the only non-trivial brackets being
\begin{subequations}\begin{align}
\big\{\bold{\varepsilon}_{ab}(x),\omega_c(x')\big\}_{\mathcal{N}}&=8\pi G\,\uepsilon_{abc}\delta^{(3)}(x,x'),\\
\big\{\bold{\eta}_{Aab}(x),\ou{\psi}{B}{c}(x')\big\}_{\mathcal{N}}&=8\pi\I\,G\,\uepsilon_{abc}\delta^B_A\delta^{(3)}_{\mathcal{N}}(x,x').
\end{align}\label{Poiss2}\end{subequations}

\subsection{Gauge symmetries}\label{3.2}
Next, we study the canonical gauge symmetries of the theory. For definiteness, we will only discuss the situation at $\varSigma\equiv\varSigma_1$, where we introduce the pre-symplectic two-form $\Omega_\varSigma$, which is derived from the second variation of the action (the exterior functional differential of the pre-symplectic potential), namely
\begin{equation}
\Omega_{\varSigma}=\bbvar{d}\Theta_{\varSigma}.\label{Omdef}
\end{equation}
The canonical gauge symmetries are now those variations $\delta$ (linearised solutions of the field equations) that define degenerate directions of the pre-symplectic potential\,---\,in other words, those variations for which
\begin{equation}
\delta\hook\Omega_{\varSigma}\equiv\Omega_\varSigma(\delta,\cdot)=0.
\end{equation}
There are three relevant kind of gauge transformations in the setup, namely diffeomorphisms that do not move the inner and outer three-boundaries, internal $SL(2,\C)$ frame rotations, and $U(1)$ rotations of the boundary spinors.\vspace{0.5em}

\noindent\emph{(i. U(1) rotations of the spinors)} First of all, let us consider the transformations \eref{complxtrafo1} of the spinors alone. At the infinitesimal level, these complexified $U(1)$ transformations define a vector field $\delta_{\lambda+\I\phi}$ on the covariant phase space, namely the infinitesimal field variation
\begin{subequations}\begin{align}
\delta_{\lambda+\I\phi}[\ell^A]&=-\frac{\lambda+\I\phi}{2}\ell^A,\\
\delta_{\lambda+\I\phi}[\bold{\eta}_A]&=+\frac{\lambda+\I\phi}{2}\bold{\eta}_A,
\end{align}\label{U1trafo}\end{subequations}
for local gauge parameters $\lambda, \varphi:\mathcal{N}\rightarrow\R$. It is now immediate to see that
\begin{equation}
\delta_{\lambda+\I\phi}\hook\Omega_\varSigma=\frac{\I}{16\pi G}\int_{\mathcal{C}}\left((\lambda+\I\phi)\bbvar{d}(\bold{\eta}_A\ell^A)-\CC\right)=-\frac{1}{8\pi G}\int_{\mathcal{C}}\lambda\bbvar{d}\bold{\varepsilon},\label{Qdef1}
\end{equation}
where we used the definition \eref{areadef} of the area two-form in terms of the boundary spinors. Equation \eref{Qdef1} shows that the infinitesimal $U(1)$ transformations $\delta_{\I\phi}$ generate local gauge symmetries of the theory. Dilatations $\delta_{\lambda}$ are, on the other hand, Hamiltonian motions (in contrast to gauge symmetries), and they are generated by the charges
\begin{equation}
Q_\lambda[\mathcal{C}]=\frac{1}{8\pi G}\int_{\mathcal{C}}\lambda\bold{\varepsilon},\label{Arham}
\end{equation}
whose Hamiltonian vector field $\mathfrak{X}_{Q_\lambda}=\{Q_\lambda,\cdot\}$ clearly satisfies the Hamilton equations
\begin{equation}
\Omega_\varSigma(\mathfrak{X}_{Q_\lambda},\cdot)=-\bbvar{d}Q_\lambda[\mathcal{C}].
\end{equation}
\vspace{0.5em}

\noindent\emph{(ii. local frame rotations)}
Next, we consider internal frame rotations. Given a local gauge parameter $\ou{\Lambda}{A}{B}\in\mathfrak{sl}(2,\C)$, we define infinitesimal Lorentz transformations, namely in the bulk
\begin{subequations}\begin{align}
\delta_\Lambda\Sigma_{AB}&=-2\ou{\Lambda}{C}{(A}\Sigma_{B)C}\\
\delta_\Lambda A^{AB}&=-\nabla\Lambda^{AB},
\end{align}\label{SLtrafo1}\end{subequations}
at the internal boundary
\begin{subequations}\begin{align}
\delta_\Lambda\ell^A&=\ou{\Lambda}{A}{B}\ell^B,\\
\delta_\Lambda\bold{\eta}^A&=\ou{\Lambda}{A}{B}\bold{\eta}^{B},
\end{align}\label{SLtrafo2}\end{subequations}
and at the outer time-like cylinder
\begin{equation}
\delta_\Lambda z^{AA'}=\ou{\Lambda}{A}{B}z^{BA'}+\ou{\bar\Lambda}{A'}{B'}z^{AB'},\label{SLtrafo3}
\end{equation}
where we allowed Lorentz transformations of the internal and spacelike normal vector $z^\alpha$ as well. A short calculations gives
\begin{align}
\delta_\Lambda\hook\Omega_\varSigma=0.
\end{align}
This can be done more explicitly. Taking into account the torsionless condition \eref{trsnless}, the glueing conditions \eref{glu} and the orientation of the boundary $\partial\varSigma=\mathcal{C}^{-1}_{in}\cup\mathcal{C}_{out}$, we find, indeed,
\begin{align}
\delta_\lambda\hook\Omega_\varSigma=&+\frac{\I}{8\pi G}\int_{\varSigma}\big(\bbvar{d}\Sigma_{AB}\wedge \nabla\Lambda^{AB}-2\ou{\Lambda}{C}{A}\Sigma_{CB}\wedge\bbvar{d}A^{AB}-\CC\big)+\nonumber\\
&+\frac{\I}{8\pi G}\int_{\mathcal{C}^{in}}\left(\Lambda^{AB}\bbvar{d}(\bold{\eta}_A\ell_B)-\CC\right)+\frac{1}{8\pi G}\int_{\mathcal{C}^{out}}\bbvar{d}(\ast\Sigma_{\alpha\beta}z^\alpha z_\mu)\Lambda^{\beta \mu}=\nonumber\\
=&+\frac{\I}{8\pi G}\int_{\varSigma}\nabla\big(\Lambda^{AB}\bbvar{d}\Sigma_{AB})
+\frac{\I}{8\pi G}\int_{\mathcal{C}^{in}}\left(\Lambda^{AB}\bbvar{d}(\bold{\eta}_A\ell_B)-\CC\right)+\nonumber\\&-\frac{1}{16\pi G}\int_{\mathcal{C}^{out}}\bbvar{d}(\ast\Sigma_{\alpha\beta})\Lambda^{\alpha\beta}=0.
\end{align}

In summary, all local frame rotations are gauge symmetries of the system. Clearly, this is all not very surprising: In the absence of fermions, we can always integrate out internal gauge transformations by simply going to the metric formulation. Therefore, no additional charges coming from internal frame rotations are to be expected.
\vspace{0.5em}

\noindent\emph{(iii. diffeomorphisms)} Working with gauge connection variables, we first lift the diffeomorphisms from the base manifold into the bundle. The horizontal lift $\mathcal{L}_\xi$ of the ordinary Lie derivative $L_\xi(\cdot)=\di\xi\hook(\cdot)+\xi\hook\di(\cdot)$ from the base manifold $\mathcal{M}$ into the spin bundle is then defined as follows
\begin{subequations}\begin{align}
\mathcal{L}_\xi\ou{\Sigma}{A}{B}&=\nabla(\xi\hook\ou{\Sigma}{A}{B})+\xi\hook \nabla\Sigma=\nabla(\xi\hook\ou{\Sigma}{A}{B}),\\
\mathcal{L}_\xi\ou{A}{A}{B}&=\xi\hook\ou{F}{A}{B},
\end{align}\label{gaugedlie}\end{subequations}
where $\nabla$ denotes the exterior covariant derivative with respect to the torsionless connection in the bulk, as in e.g.\ \eref{Eeq} above. In the same way, the Lie derivative extends to the fields on the two  boundaries, namely
\begin{subequations}\begin{align}
\mathcal{L}_\xi\ell^A&=\xi\hook D\ell^A=\xi^a D_a\ell^A,\\
\mathcal{L}_\xi\bold{\eta}_{A}&=D(\xi\hook\bold{\eta}_{A})+\xi\hook D\bold{\eta}_{A},\\
\mathcal{L}_\xi z^\alpha&=\xi\hook D z^\alpha=\xi^aD_a z^\alpha,\label{Liezet}\end{align}%
\label{Liedefs}%
\end{subequations}%
with $D_a$ denoting the pull-back \eref{Ddef} of the covariant derivative $\nabla_a$ to either the internal null boundary $\mathcal{N}$ or the outer cylinder approaching spacelike infinity. Notice that the vector field $\xi^a\in T\mathcal{M}$ has to be tangential to the two boundary components
\begin{equation}
\xi^a\big|_{\mathcal{N}}\in T\mathcal{N},\quad \xi^a\big|_{\mathcal{B}}\in T\mathcal{B},
\end{equation}
for otherwise it would not make sense to speak about the Lie derivative of boundary fields, which is well-defined only for vectors $\xi^a$ that are tangential to the boundary. 

We now compute the interior product of the field variation $\delta_\xi(\cdot)=\mathcal{L}_\xi(\cdot)$ with the pre-symplectic two-form \eref{Omdef}, namely
\begin{equation}
\Omega_\varSigma(\delta,\mathcal{L}_\xi)=\delta\left[\Theta_\varSigma(\mathcal{L}_\xi)\right]-\mathcal{L}_\xi\left[\Theta_\varSigma(\delta)\right]-\Theta_\varSigma\big([\delta,\mathcal{L}_\xi]\big),
\end{equation}
where $\delta\equiv(\delta\ou{A}{A}{B},\delta\Sigma_{AB})$ is a linearised solution of the field equations \eref{Eeq} (a tangent vector in the covariant phase space). Any such tangent vector satisfies the identity
\begin{equation}
\int_\varSigma\left(\delta\Sigma_{AB}\wedge\xi\hook F^{AB}-(\xi\hook\Sigma_{AB})\wedge\delta F^{AB}\right)=0, 
\end{equation}
which is a consequence of the Einstein equations. From there, one proceeds to show
\begin{align}\nonumber
\Omega_\varSigma&(\delta,\mathcal{L}_\xi)=\\\nonumber
=&+\frac{\I}{8\pi G}\int_{\mathcal{C}_{in}}\!\!\left(\delta\bold{\eta}_A\mathcal{L}_\xi\ell^A-\mathcal{L}_\xi\bold{\eta}_A\delta\ell^A
+(\xi\hook\Sigma_{AB})\wedge\delta A^{AB}-\CC\right)+\\\nonumber
&-\frac{1}{8\pi G}\int_{\mathcal{C}_{out}}\!\!\Big(\delta(\ast\Sigma_{\alpha\beta})z^\alpha\mathcal{L}_\xi z^\beta-\mathcal{L}_\xi(\ast\Sigma_{\alpha\beta})\wedge z^\alpha\delta z^\beta+\\
&\hspace{14em}+\frac{1}{2}(\xi\hook\ast\!\Sigma_{\alpha\beta})\wedge{\delta A}^{\alpha\beta}\Big).\label{Hvar}
\end{align}%
If $\xi^a$ tangential to the two corners, the expression simplifies further and turns into a total functional differential,
\begin{align}\nonumber
\Omega_\varSigma(\delta,\mathcal{L}_\xi)=-\delta J_{\xi}[\mathcal{C}_{out}]+\delta J_\xi[\mathcal{C}_{in}].
\end{align}
Where we introduced the diffeomorphism charges
\begin{subequations}\begin{align}
J_{\xi}[\mathcal{C}_{in}]&=\frac{\I}{8\pi G}\int_{\mathcal{C}_{in}}\left(\bold{\eta}_A\mathcal{L}_\xi\ell^A-\CC\right),\label{surfchrge}\\
J_{\xi}[\mathcal{C}_{out}]&=\frac{1}{8\pi G}\int_{\mathcal{C}_{out}}\ast\Sigma_{\alpha\beta}\,z^\alpha\mathcal{L}_\xi z^\beta.\label{ADMcharge}
\end{align}\label{diffchargs}\end{subequations}
at the two corners. The integrability of $J_\xi[\mathcal{C}]$ follows from \eref{Hvar} by Stokes's theorem and once commuting $\delta$ with $\mathcal{L}_\xi$, which returns an infinitesimal $SL(2,\C)$ gauge transformation
\begin{align}\nonumber
[\delta,\mathcal{L}_\xi]&A^{AB}=-\nabla\xi\hook\delta A^{AB},\hspace{-2em}&[\delta,\mathcal{L}_\xi]&\ell^A=(\xi\hook\delta\ou{A}{A}{B})\ell^B,\\\nonumber
[\delta,\mathcal{L}_\xi]&\Sigma_{AB}=-2(\xi\hook\delta\ou{A}{C}{(A})\Sigma_{B)C},\hspace{-2em}&[\delta,\mathcal{L}_\xi]&\bold{\eta}_A=-(\xi\hook\delta\ou{A}{B}{A})\bold{\eta}_B,\\
[\delta,\mathcal{L}_\xi]&z^\alpha=(\xi\hook \delta\ou{A}{\alpha}{\beta})z^\beta,&
\end{align}
which is a degenerate direction of the pre-symplectic two-form $\Omega_\varSigma$. 

The diffeomorphism charges \eref{diffchargs} are defined for all vector fields that are tangential to the corners
\begin{equation}
\xi^a\big|_{\mathcal{C}^{out}_{in}}\in T\mathcal{C}^{out}_{in}.
\end{equation}
These charges are nothing but the canonical Noether charges, and they assume the canonical form of an interior product of the field variation $\delta_\xi(\cdot)=\mathcal{L}_\xi$ with the covariant symplectic potential \eref{actnpot}. 

In the limit\footnote{Introducing asymptotically inertial coordinates $\{t,x^i\}$, we define the outer boundary as a $\rho=\sqrt{\eta_{\alpha\beta}x^\alpha x^\beta}=\mathrm{const}.$ hyperbolic cylinder and the outer corner $\mathcal{C}_{out}$ as a $t=\mathrm{const}.$ cross-section thereof. The $\rho\rightarrow\infty$ limit of $J_\xi[\mathcal{C}_{out}]$ returns the ADM angular momentum for an asymptotic rotation such as $\xi^a_{(i)}=\uo{\epsilon}{ij}{k}x^j\partial^a_k+\mathcal{O}_-(\rho^0)$, where the order $\rho^0$ subleading term $\mathcal{O}_-(\rho^0)$ is parity odd, and the usual falloff and parity condition for metric and extrinsic curvature are satisfied \cite{ReggeTeitelboim}.\label{asymptcsnote}} where the outer cylinder approaches spacelike infinity, the integral \eref{ADMcharge} returns the ADM angular momentum provided $\xi^a\in T\mathcal{C}_{out}$ is an asymptotic rotation. This can be seen as follows: Let $z^\alpha=\ou{e}{\alpha}{a}z^a$ be the internal and outwardly oriented normal to the outer cylinder $\mathcal{B}$. Choose a foliation such that the partial Cauchy hypersurface $\varSigma$ intersects the outer cylinder $\mathcal{B}$ orthogonally, and let $n^\alpha=\ou{e}{\alpha}{a}n^a$ be the internal and future oriented normal vector to $\varSigma$. We now have a normalised dyad $(n^\alpha,z^\alpha):n_\alpha z^\alpha=0$ at the corner $\mathcal{C}=\mathcal{B}\cap\varSigma$. The canonical surface area element at this outer corner is given by $d^2v=\ast\Sigma_{\alpha\beta}n^\alpha z^\beta$. For any $\xi^a\in T\mathcal{C}_{out}$, we have $\xi^an_a=\xi^az_a=0$, such that
\begin{align}\nonumber
\int_{\mathcal{C}}\ast\Sigma_{\alpha\beta}\,z^\alpha\mathcal{L}_\xi z^\beta&=\int_{\mathcal{C}}d^2v\,n_\alpha\mathcal{L}_\xi z^\alpha=-\int_{\mathcal{C}}d^2v\,z^\alpha\mathcal{L}_\xi n_\alpha=\\
&=-\int_{\mathcal{C}}d^2v\,z^a\xi^b\nabla_bn_a=-\int_{\mathcal{C}}d^2v\,z^a(K_{ab}-h_{ab}K)\xi^b,\label{ADmdens}
\end{align}
where $K_{ab}$ denotes the extrinsic curvature of $\varSigma$, and the gauge covariant Lie derivative $\mathcal{L}_\xi z^\alpha$ is defined as in \eref{Liezet}. The canonical ADM momentum tensor density is $\tilde{\pi}^{ab}=1/(16\pi G)\,d^3v\,(K^{ab}-h^{ab}K)$, such that equation \eref{ADMcharge} does, indeed, return the expression for the ADM angular momentum.

The charges \eref{surfchrge} and \eref{ADMcharge} are defined for vector fields $\xi^a$ that are \emph{tangential} to the outer (inner) corners. What happens for diffeomorphisms that move the corner forward in time? The question will be most interesting at the inner null surface, and we will study this case in the next section. At the outer cylinder, on the other hand, the answer is clear and all very well known: Asymptotic time translations $t^a$ are generated by the ADM four-momentum $P_\alpha$, whose variation  is given by\footnote{See e.g.\ \cite{Corichi:2015cqa,Ashtekar:2008jw} for a recent analysis using connection variables.}
\begin{equation}
\delta P_\alpha t^\alpha=\frac{1}{16\pi G}\lim_{\rho\rightarrow\infty}\int_{\mathcal{C}_{\rho}}(t\hook\ast\!\Sigma_{\alpha\beta})\wedge\delta A^{\alpha\beta}.\label{varADM}
\end{equation}
For the completeness of the paper, we sketch the proof in \hyperref[appdxB]{appendix B}. Consider then such an asymptotic time translation $t^a|_{\mathcal{C}_{out}}\in T\mathcal{B}$. Suppose, for simplicity, that $t^a$ vanishes at the inner null surface. Going back to \eref{Hvar} and taking into account the parity and falloff conditions for metric and connection \cite{ReggeTeitelboim}, one can then show that all but the last term of equation \eref{Hvar} vanish for $\rho\rightarrow\infty$. This last term has a limit for $\rho\rightarrow\infty$,  and this limit can be written as the total variation of the ADM energy (see \hyperref[appdxB]{appendix B}), such that asymptotic time translations $\mathcal{L}_t$ are indeed integrable. In other words,
\begin{equation}
\Omega_\varSigma(\delta,\mathcal{L}_t)=-\frac{1}{16\pi G}\lim_{\rho\rightarrow\infty}\int_{\mathcal{C}_{\rho}}(t\hook\ast\!\Sigma_{\alpha\beta})\wedge\delta A^{\alpha\beta}=-\delta P_\alpha t^\alpha=\delta E_t.\label{ADMchrgdef}
\end{equation}


Before we proceed to the next section, I briefly summarise the main message of this section. We identified three kind of gauge transformations: Local $U(1)$ transformations $\delta_{\I\phi}$ of the boundary spinors 
\eref{U1trafo}, local $SL(2,\C)$ frame rotations $\delta_{\Lambda}$ in the bulk and boundary (\ref{SLtrafo1}, \ref{SLtrafo2}, \ref{SLtrafo3}), and diffeomorphisms $\delta_\xi=\mathcal{L}_\xi$ that preserve the two corners bounding the bulk. On the other hand, there are then also those diffeomorphisms that do not preserve the corners $\mathcal{C}_{in}$ and  $\mathcal{C}_{out}$, but drag them further along the boundary. In the absence of additional structures (exact or asymptotic Killing vectors), such diffeomorphisms $\mathcal{L}_\xi$ will not define gauge symmetries. They will define genuine Hamiltonian motions, and for certain such vector fields $\xi^a$, we found the boundary charges that generate those motions. We considered vector fields $\xi^a$ that can be written as the sum
\begin{equation}
\xi^a=t^a+\varphi^a,\label{xicomp}
\end{equation}
of an asymptotic time translation $t^a$ that vanishes at the inner corner and a vector field $\varphi^a$ that is tangential to both corners. It can be then inferred from equations \eref{ADMchrgdef} and \eref{diffchargs} that such a vector field  defines a Hamiltonian flow on phase space, which is generated by a charge, whose on-shell value is given by
\begin{equation}
H_\xi=E_t-J_\varphi^\infty+J_\varphi[\mathcal{C}_{in}],\label{SumCharges}
\end{equation}
consisting of the ADM energy \eref{ADMchrgdef} at infinity and the angular moments\footnote{$J_\varphi^\infty$ denotes the $\rho\rightarrow\infty$ limit of the ADM charge \eref{ADMcharge}, which returns the angular momentum for an asymptotic rotation $\varphi^a$.} \eref{diffchargs}. Besides these diffeomorphism charges, we also introduced in \eref{Arham} the  \emph{area Hamiltonian} $Q_\lambda[\mathcal{C}_{in}]$ on the inner null surface, that generates local dilatations of the boundary spinors, namely $\ell^A\rightarrow\E^{-\lambda/2}\ell^A$ and $\bold{\eta}_A\rightarrow\E^{+\lambda/2}\bold{\eta}_A$. Such a charge plays an important role in loop quantum gravity and the physics of black holes, c.f.\ 
\cite{Carlip:1993sa,Bianchi:2012ev,FGPfirstlaw}.
\subsection{Quasi-local first law}\label{sec3.3}
So far, the vector field $\xi^a$ was assumed to be tangential to the inner corner, and it was shown that the resulting diffeomorphism $\exp(\xi)$ is generated by a Hamiltonian, whose on-shell value is given by the diffeomorphism charges \eref{SumCharges}. We now wish to consider the other extreme, where $\xi^a$ no longer preserves the inner corner, but drags it along the null generators. 
For definiteness, we assume that $\xi^a$ vanishes at infinity, while at the inner null boundary, it is defined implicitly in terms of the boundary spinor $\ell^A$ and the inverse tetrad, namely through
\begin{equation}
\xi^a_{(\alpha)}\big|_{\mathcal{N}}=\alpha\,\I\,\uo{e}{AA'}{a}\ell^A\bar{\ell}^{A'},\label{xivdef}
\end{equation}where $\alpha:\mathcal{N}\rightarrow\R$ is a lapse function. 
The tetrad $\ou{e}{AA'}{a}$ and the boundary spinors are coordinates on phase space. This implies that the vector field $\xi^a_{(\alpha)}$, which is parallel to the null generators of the boundary, 
depends (as a function) on the points $z\in\mathcal{N}$ and (as a functional) on the field configuration $\mathcal{p}=(\ou{\Sigma}{A}{Bab},\ou{A}{A}{Ba},\bold{\eta}_{Aab},\b\ell^{A'})$ on the covariant phase space of the theory. The vector field \eref{xivdef} defines, therefore, a field dependent\footnote{The geometry of such field dependent transformations was studied carefully in a recent paper by Aldo Riello and Henrique Gomes \cite{Gomes:2016mwl}.} 
phase space variation $\delta_{(\alpha)}=\mathcal{L}_{\xi_{(\alpha)}}$.  

Above, we saw that asymptotic Poincaré transformations $\delta_{t+\varphi}=\mathcal{L}_{t+\varphi}$ that preserve the inner corner are integrable\,---\,the on-shell value of the generator being a sum of the ADM energy \eref{ADMchrgdef} and the generalised angular moments \eref{diffchargs}. We now want to see under which conditions the field variation $\delta_{(\alpha)}=\mathcal{L}_{\xi_{(\alpha)}}$ that drags the inner corner along the null generators is Hamiltonian as well\,---\,under which conditions there exists a Hamiltonian $H_{(\alpha)}[\mathcal{C}_{in}]$ on the covariant phase space, such that
\begin{equation}
\Omega_\varSigma(\delta,\delta_{{(\alpha)}})=-\delta H_{\xi_{(\alpha)}}[\mathcal{C}_{in}].
\end{equation}
Equation \eref{Hvar} determines the interior product $\mathcal{L}_\xi\hook\Omega$ of an arbitrary such field variation $\mathcal{L}_\xi$ with the pre-symplectic two-form \eref{Omdef}. The desired integrability condition for $H_{(\alpha)}[\mathcal{C}_{in}]$ is therefore given by
\begin{align}\nonumber
\delta H_{(\alpha)}[\mathcal{C}_{in}]=-\frac{\I}{8\pi G}\int_{\mathcal{C}_{in}}\!\!&\Big(\delta\bold{\eta}_A\mathcal{L}_{\xi_{(\alpha)}}\ell^A-\mathcal{L}_{\xi_{(\alpha)}}\bold{\eta}_A\delta\ell^A
+\\
&\qquad+(\xi_{(\alpha)}\hook\Sigma_{AB})\wedge\delta A^{AB}-\CC\Big).\label{Hvertvar}
\end{align}
To identify the geometric significance of the various terms in the expression, we separate physical relevant variations from those directions in phase space that define internal gauge transformations. This amounts to splitting the field variation $\delta$ into \emph{horizontal} and \emph{vertical} components, thereby implicitly introducing a connection  on field space, such as in e.g.\ \cite{Gomes:2016mwl}. A natural such decomposition
\begin{equation}
\delta=\delta_H+\delta_V,\label{HVcomps}
\end{equation}
can be introduced as follows: Consider the pull-back $\varphi_{\mathcal{C}_{in}}^\ast:T^\ast\mathcal{N}\rightarrow T^\ast\mathcal{C}_{in}$ from the null surface to the two-dimensional and spatial cross section $\mathcal{C}_{in}\subset\mathcal{N}$. To clarify our notation, we denote the corresponding tensor indices on $T^\ast\mathcal{C}_{in}$ by $\ucheck{a},\ucheck{b},\dots$, thus writing, e.g.
\begin{equation}
(\varphi_{\mathcal{C}_{in}}^\ast\bold{\eta}{}_A){}_{ab}\equiv\bold{\eta}{}_{A\underaccent{\check}{a}\underaccent{\check}{b}}.\label{pullbckcrnr}
\end{equation}
We can then always achieve a gauge transformation $k_A\rightarrow k_A+\zeta\ell_A$, as in \eref{gaugespin} above, such that the dual spinor $k_A$, which we used for the decomposition \eref{mudecomp} of $\bold{\eta}_{Aab}$ into component functions $\bold{\mu}_{ab}$ and $\bold{\varepsilon}_{ab}$, is implicitly defined by the pull-back
\begin{equation}
\bold{\eta}_{A\underaccent{\check}{a}\underaccent{\check}{b}}=\I k_A\bold{\varepsilon}_{\ucheck{a}\ucheck{b}}\label{etapullbck}
\end{equation}
 of the area two-form \eref{areadef} to the two-dimensional corner. Given the boundary spinor $(\bold{\eta}_{Aab},\bar\ell^{A'})$ on $\mathcal{N}$, we now have a canonical prescription to introduce a normalised spin dyad $(k_A,\ell_A):k_A\ell^A=1$ at the inner corner. Any two such normalised spin dyads are related by an $SL(2,\C)$ gauge transformation that sends one basis into the other. This implies that for every infinitesimal field variation $\delta$, there must exists some specific $\mathfrak{sl}(2,\C)$ Lie algebra element $\ou{[\Lambda_\delta]}{A}{B}$, such that 
\begin{equation}
\delta[\ell^A]=\ou{[\Lambda_\delta]}{A}{B}\ell^B,\quad \delta[k^A]=\ou{[\Lambda_\delta]}{A}{B}k^B.\label{gaugeparam}
\end{equation}
We can then identify the vertical components \eref{HVcomps} of the field variation $\delta$ on the null surface with an infinitesimal $SL(2,\C)$ gauge transformation $\delta_\Lambda$ (\ref{SLtrafo1}, \ref{SLtrafo2}) for a gauge parameter $\ou{\Lambda}{A}{B}$ that is implicitly defined by \eref{gaugeparam}. This is the same as to say that the horizontal component of the field variation $\delta$ annihilates the spinors
\begin{equation}
\delta_H[\ell^A]=\delta_H[k^A]=0.
\end{equation}
The vertical component, on the other hand, is an internal gauge transformation, whose Hamiltonian vector field defines a degenerate direction of the pre-symplectic two-form, such that
\begin{equation}
\Omega_\varSigma(\delta,\delta_{\xi_{(\alpha)}})=\Omega_\varSigma(\delta_H,\delta_{\xi_{(\alpha)}}).
\end{equation}
Going back to the integrability condition \eref{Hvertvar} for the Hamiltonian, we can conclude, therefore, that
\begin{align}\nonumber
\delta H_{(\alpha)}[\mathcal{C}_{in}]=-\frac{\I}{8\pi G}\int_{\mathcal{C}_{in}}\!\!\Big(k_A\mathcal{L}_{\xi_{(\alpha)}}&\ell^A\delta_H{\boldsymbol{\varepsilon}}+\\
&+(\xi_{(\alpha)}\hook\Sigma_{AB})\wedge\delta_HA^{AB}-\CC\Big).\label{Hvertvar2}
\end{align}
This equation can be simplified further. The vector field $\xi^a_{(\alpha)}$ lies inside the null boundary, in evaluating $\xi_{(\alpha)}\hook\Sigma_{AB}$ on $\mathcal{N}$, we can use, therefore, the glueing condition \eref{glu} that determines the pull-back of $\Sigma_{AB}$ in terms of the boundary spinors $\ell_A$ and $\bold{\eta}_{Aab}$. This implies 
\begin{equation}
\varphi^{\ast}_{\mathcal{N}}(\xi_{(\alpha)}\hook\Sigma_{AB})=\xi_{(\alpha)}\hook\bold{\eta}_{(A}\ell_{B)}=-\alpha \ell_A\ell_B \bar{m},
\end{equation}
and hence also
\begin{equation}
\delta H_{(\alpha)}[\mathcal{C}_{in}]=-\frac{\I}{8\pi G}\int_{\mathcal{C}_{in}}\!\!\Big(k_A\mathcal{L}_{\xi_{(\alpha)}}\ell^A\delta_H{\boldsymbol{\varepsilon}}+\alpha\,\bar{m}\wedge\delta_H(\ell_AD\ell^A)-\CC\Big).\label{Hvertvar3}
\end{equation}
All of these terms have a straightforward geometric interpretation. The surface gravity $\kappa_{(\ell)}$ is defined as the acceleration of $\ell^a$, namely
\begin{equation}
\ell^b\nabla_b\ell^a=\kappa_{(\ell)}\ell^a,
\end{equation}
in terms of the boundary spinors, this becomes
\begin{equation}
\alpha\kappa_{(\ell)}=k_A\mathcal{L}_{\xi_{(\alpha)}}\ell^A+\CC
\end{equation}
In the same way, shear $\sigma_{(\ell)}$ and expansion $\vartheta_{(\ell)}$ of the null congruence generating the null surface $\mathcal{N}$ can be written 
 in terms of the spinors and their derivatives. Following Penrose's conventions \cite{penroserindler}, we have
\begin{subalign}
\sigma_{(\ell)}&=-\ell_Am^{\ucheck{a}}D_{\ucheck{a}}\ell^A=m^{\ucheck{a}}m^{\ucheck{b}}\nabla_{\ucheck{a}}\ell_{\ucheck{b}},\\
\frac{1}{2}\vartheta_{(\ell)}&=-\ell_A\bar{m}^{\ucheck{a}}D_{\ucheck{a}}\ell^A=\frac{1}{2}q^{\ucheck{a}\ucheck{b}}\nabla_{\ucheck{a}}\ell_{\ucheck{b}},
\end{subalign}
where $m^{\ucheck{a}}$ is built from the pull back $T^\ast\mathcal{N}\rightarrow T^\ast\mathcal{C}_{in}$ of $m_a$ by  raising the index with the inverse\footnote{On a null surface $\mathcal{N}$, the pull-back $\varphi^\ast_{\mathcal{N}}g_{ab}=q_{ab}=2m_{(a}\bar{m}_{b)}$ defines a degenerate signature $(0$$+$$+)$ metric, which has no inverse. If we go down a dimension further, and consider the pull-back $q_{\ucheck{a}\ucheck{b}}$ of $q_{ab}$ to the two-dimensional spatial cross-section $\mathcal{C}_{in}\subset\mathcal{N}$, an inverse metric $q^{\ucheck{a}\ucheck{b}}\in T\mathcal{C}_{in}\otimes T\mathcal{C}_{in}$ exists again.} and two-dimensional signature $(+$$+)$ metric $q^{\ucheck{a}\ucheck{b}}$ on the inner corner,
\begin{equation}
q^{\ucheck{a}\ucheck{c}}q_{\ucheck{b}\ucheck{c}}=\delta^{\ucheck{a}}_{\ucheck{b}}.
\end{equation}
Having introduced shear, expansion and surface gravity, we can bring the integrability condition \eref{Hvertvar3} into the following form
\begin{align}\nonumber
\delta H_{(\alpha)}[\mathcal{C}_{in}]=\frac{1}{8\pi G}\int_{\mathcal{C}_{in}}\alpha
\left(\kappa_{(\ell)}\delta_H\right.&\bold{\varepsilon}-\frac{1}{2}\vartheta_{(\ell)}\delta_H\bold{\varepsilon}-\bold{\varepsilon}\delta_H\vartheta_{(\ell)}+\\
&\left.+\I\sigma_{(\ell)}\bar{m}\wedge\delta_H\bar{m}-\I\bar\sigma_{(\ell)}{m}\wedge\delta_H{m}\right).\label{Hvertvar4}
\end{align}
Next, we define the following tensor
\begin{equation}
\Sigma_{(\ell)}^{\ucheck{a}\ucheck{b}}=\bar\sigma_{(\ell)}m^{\ucheck{a}}m^{\ucheck{b}}+\sigma_{(\ell)}\bar{m}^{\ucheck{a}}\bar{m}^{\ucheck{b}}+\frac{1}{2}q^{\ucheck{a}\ucheck{b}}\vartheta_{(\ell)}\in T\mathcal{C}_{in}\otimes T\mathcal{C}_{in}.
\end{equation}
In \eref{Hvertvar4}, all horizontal variations $\delta_H$ act on fields that are already $SL(2,\C)$ gauge invariant. This means that we can replace the horizontal derivative $\delta_H$ by the ordinary variations \eref{HVcomps}, such that e.g.\
\begin{equation}
\delta_H\bold{\varepsilon}_{\ucheck{a}\ucheck{b}}=\delta\bold{\varepsilon}_{\ucheck{a}\ucheck{b}}=\frac{1}{2}\bold{\varepsilon}_{\ucheck{a}\ucheck{b}}q^{\ucheck{c}\ucheck{d}}\delta q_{\ucheck{c}\ucheck{d}}.\label{Hepsvar}
\end{equation}
Bringing this together with \eref{Hvertvar4}, we finally get
\begin{align}\nonumber
\delta H_{(\alpha)}[\mathcal{C}_{in}]&\stackrel{!}{=}-\Omega_\varSigma(\delta,\mathcal{L}_{\xi_{(\alpha)}})=\\
&=\frac{1}{8\pi G}{\int_{\mathcal{C}_{in}}}\!\!\!\alpha
\Big(\kappa_{(\ell)}\delta\bold{\varepsilon}-\bold{\varepsilon}\delta\vartheta_{(\ell)}-\frac{1}{2}\bold{\varepsilon}\Sigma_{(\ell)}^{\ucheck{a}\ucheck{b}}\delta q_{\ucheck{a}\ucheck{b}}\Big).\label{fstlaw}
\end{align}
This is a quasi-local version of the first law, and it appears in here as an integrability condition for the Hamiltonian $H_{(\alpha)}[\mathcal{C}_{in}]$, which exists, if the quasi-local first law \eref{fstlaw} is satisfied. 
\vspace{1em}

In general, the vector field $\xi^a_{(\alpha)}|_{\mathcal{N}}=\alpha\ell^a$ that drags the inner corner along the null surface will not be integrable, and there will be no Hamiltonian, unless, however, the generalised first law \eref{fstlaw} is satisfied. When is it then satisfied? A minimal example is given by a Killing vector
\begin{equation}
\xi^a=\ell^a+t^a+\varphi^a,\label{Killngvec1}
\end{equation}
where  both $\varphi^a$ and $t^a$ are assumed to vanish at the inner null surface (with their first derivatives). At the null surface, the Killing vector $\xi^a$ is null itself and implicitly given by \eref{xivdef}. The other two components are the asymptotic time translation  $t^a$, and an asymptotic rotation $\varphi^a$. Now, $\xi^a$ is assumed to be a Killing vector, and it defines, therefore, an infinitesimal and internal Lorentz transformation for a gauge parameter $\ou{[\Phi_\xi]}{A}{B}\in \mathfrak{sl}(2,\C)$, which is given by
\begin{equation}
\nabla_a\xi_b=\Phi_{ba}=-\Phi_{ab},\quad\Phi_{AB}:=\frac{1}{2}\uo{\Sigma}{AB}{ab}\Phi_{ab},\label{Killngvec2}
\end{equation}
where $\Sigma_{ABab}$ are the components of the Pleba\'nski self-dual two-form \eref{Plebform}. At the level of the spin bundle, the gauge covariant Lie derivative \eref{gaugedlie} drags the fields in the bulk up to an internal $SL(2,\C)$ gauge transformation, for a gauge parameter $\ou{[\Phi_\xi]}{A}{B}\in\mathfrak{sl}(2,\C)$ which is given by \eref{Killngvec2}, such that
\begin{subalign}
\mathcal{L}_\xi\Sigma_{AB}&=\nabla(\xi\hook\Sigma_{AB})=-2\ou{[\Phi_\xi]}{C}{(A}\Sigma_{B)C},\\
\mathcal{L}_\xi\ou{A}{A}{B}&=\xi\hook\ou{F}{A}{B}=-\nabla\ou{[\Phi_\xi]}{A}{B}.
\end{subalign}
A Killing vector defines a symmetry (in the Hamiltonian sense) for a given and fixed field configuration $\mathcal{p}=(\Sigma_{AB},A^{AB},\ell^A,\bold{\eta}_A,z^\alpha)$, only if the infinitesimal field variation $\delta_\xi[\cdot]=\mathcal{L}_\xi(\cdot)$ is a degenerate direction of the pre-symplectic two-form,
\begin{equation}
\mathcal{L}_\xi\hook\Omega_\Sigma\big|_{\mathcal{p}}=0.
\end{equation}
This is satisfied only if $\mathcal{L}_\xi$ acts as an internal gauge symmetry on the boundary fields as well. The residual gauge transformations at the inner null boundary are $U(1)$ rotations of the spinors and $SL(2,\C)$ frame rotations (see \hyperref[sec3.2]{section 3.2}). For $\mathcal{L}_\xi$ to be a symmetry, there is, therefore, a nontrivial boundary condition to be satisfied: In addition to $\ou{[\Phi_\xi]}{A}{B}\in\mathfrak{sl}(2,\C)$ there must exist a local $U(1)$ gauge parameter $\phi_\xi$ such that
\begin{subalign}
\mathcal{L}_\xi \ell^A&=+\ou{[\Phi_\xi]}{A}{B}\ell^B+\frac{\I\phi_\xi}{2}\ell^A,\\
\mathcal{L}_\xi \bold{\eta}_A&=-\ou{[\Phi_\xi]}{B}{A}\bold{\eta}_B-\frac{\I\phi_\xi}{2}\ell^A.
\end{subalign}
In other words, the boundary fields must be Lie dragged as well (modulo gauge). 
The infinitesimal field variation $\delta_\xi=\mathcal{L}_\xi$ can be replaced, therefore, by an internal gauge transformation, which is a degenerate direction of the pre-symplectic two-form. At those specific field configurations $\mathcal{p}$ that admit a Killing vector (\ref{Killngvec1}, \ref{Killngvec2}) there exists then a Hamiltonian $H_\xi[\mathcal{C}_{in}]=E_t+J_\varphi^\infty$ such that for all field variations around $\mathcal{p}$,
\begin{equation}
0=\Omega_\varSigma(\delta,\delta_\xi)\big|_{\mathcal{p}}=\delta E_t-\delta J_\varphi^\infty-\delta H_\xi[\mathcal{C}_{in}].\label{Hamxi}
\end{equation}
 
In the same way, and more usefully, one can now also read \eref{Hamxi} as a version of the first law for generic Killing horizons. This follows from the quasi-local first law \eref{fstlaw} and the fact that $\mathcal{L}_\xi$ annihilates the pre-symplectic two-form \eref{Hamxi}, and it implies that the ADM energy $E_t$, the ADM angular momentum $J_\varphi^{\infty}$ at infinity, and the area two-form at the inner corner satisfy the following variational identity 
\begin{equation}
\bar{\delta} E_t=\bar{\delta}J_\varphi^\infty+\frac{1}{8\pi G}\int_{\mathcal{C}_{in}}
\kappa_{(\ell)}\bar{\delta}\bold{\varepsilon},\label{fstlaw2}
\end{equation}
in the class of all field variations $\bar{\delta}$ that do not change the expansion of the null surface, i.e.\ $\bar{\delta}\vartheta_{(\ell)}=0$. Notice that  it was nowhere required that the inner corner is a bifurcation surface where $\xi^a$ would vanish, which is often assumed in deriving the first law for black hole mechanics, such as in \cite{WaldBHbook}. 







\section{Conclusion, relevance for quantum gravity}
\subsection{Summary}
We first summarise the paper, then discuss its relevance for non-\-per\-tur\-ba\-tive and background independent quantum gravity.
Our main motivation was to develop an $SL(2,\C)$ gauge covariant Hamiltonian formalism for general relativity in the presence of inner null boundaries. First of all (\hyperref[sec2]{section 2}), we studied the basic kinematical structures of the spin bundle over a null surface. We found that the entire intrinsic geometry\footnote{The signature $(0$$+$$+)$ metric $q_{ab}$ and its degenerate null direction $[\ell^a]$.} of a three-dimensional null boundary can be described and reconstructed from the bispinor
\begin{equation}
Z^{\mathfrak{a}}=\begin{pmatrix}\ell^A\\\bold{\bar\eta}_{A'ab}\end{pmatrix},
\end{equation}
consisting of a spinor-valued two-form $\bold{\bar\eta}_{A'ab}$, and a spinor-valued $0$-form $\ell^A$ both intrinsic to $\mathcal{N}$. The bispinor $Z^{\mathfrak{a}}$ captures the entire geometry: The spin $(0,0)$ component $\bold{\varepsilon}_{ab}=-\I\bold{\eta}_{Aab}\ell^A$ defines the canonical area two-form \eref{areadef} of the boundary, the spin $(\tfrac{1}{2},\tfrac{1}{2})$ vector component $\I\ell^A\bar{\ell}^{A'}$ defines the internal direction of the null generators of the null surface, and the spin $(1,0)$ tensor component $\ell{}_{(A}\bold{\eta}_{B)ab}$ defines the pull-back of the Pleba\'nski two-form to the boundary. Contracting $Z^{\mathfrak{a}}$ with the dual spinor $\bar{Z}_{\mathfrak{a}}$ (see \eref{bndryspin}), one obtains an $SL(2,\C)$ gauge invariant scalar
\begin{equation}
\bar{Z}_{\mathfrak{a}}Z^{\mathfrak{a}},
\end{equation}
the vanishing of which imposes the reality condition \eref{realcond} for the canonical area two-form. 

We then saw in \hyperref[sec2.2]{section 2.2} how to describe the extrinsic curvature of the null surface as an equivalence class \eref{eqidef} of $SL(2,\C)$ gauge connections on the boundary. This equivalence class can be parameterised by a complexified $U(1)$ boundary connection $\omega_a$ and a spinor-valued one-form $\ou{\psi}{A}{a}$ (both intrinsic to the boundary). After having introduced both the intrinsic and the extrinsic geometry in terms of the boundary spinors, we found an $SL(2,\C)$ gauge invariant boundary term at the inner null surface (\hyperref[sec2.3]{section 2.3}). The resulting boundary action\footnote{Looking at the action, one may conclude $\ou{\psi}{A}{a}$ is some kind of Rarita\,--\,Schwinger field, because the boundary action is clearly invariant under shifts $\ell^A\rightarrow\ell^A+\varepsilon^A$, and $\ou{\psi}{A}{a}\rightarrow\ou{\psi}{A}{a}+(D_a-\omega_a)\varepsilon^A$ for some gauge parameter $\varepsilon^A$, but this pseudo shift-symmetry is broken by the coupling to the bulk: Variation of the bulk plus boundary action \eref{fullactn} yields the glueing constraint \eref{glu}, as well as the reality \eref{realcond2} and torsionless \eref{threetrsn} conditions, which are not preserved for generic $\varepsilon^A$.} 
\begin{equation}
S_{\mathcal{N}}[A|\bold{\eta},\ell|\omega,\psi]=\frac{\I}{8\pi G}\int_{\mathcal{N}}\big(\bold{\eta}_A\wedge \left(D-\omega\right)\ell^A-\bold{\eta}_A\wedge \psi^A-\CC\big)\label{actnfinal}
\end{equation}
depends on all fields previously introduced, with $D_a=\partial_a+[A_a,\cdot]$ denoting the exterior covariant derivative. The boundary conditions are that $\omega_a$ and $\ou{\psi}{A}{a}$ are fixed in the variational problem, and thus treated as \emph{external potentials} (in addition $\ell^A$ is held fixed at the boundary of the boundary, which are the inner corners $(\mathcal{C}_0)^{-1}\cup\mathcal{C}_1=\partial\mathcal{N}$, see \hyperref[fig1]{figure 1} for an illustration). The entire bulk plus boundary action yields the Einstein equations \eref{Eeq} in the bulk and the boundary equations of motion \eref{spineoms} for the boundary spinors along the null surface. 

Next (\hyperref[sec3]{section 3}), we introduced the covariant phase space of the theory. The pre-symplectic potential \eref{actnpot} at the partial Cauchy surface $\varSigma$ picks up a corner term at the intersection $\mathcal{C}=\varSigma\cap\mathcal{N}$ with the inner null surface. The new canonical variables at the inner corner are the spinor $\ell^A$ (which defines the null direction), and the pull-back to the inner corner of the spinor-valued two-form $\bold{\eta}_{Aab}$ (defining the area element). The canonical Poisson commutation relations are given in \eref{Poiss1}. Along the inner null surface, we introduced canonical commutation relations as well: The symplectic structure is determined by \eref{nullpot} and \eref{Poiss2}. The complexified $U(1)$ connection $\omega_a$ is the momentum conjugate to the canonical area two-form $\bold{\varepsilon}_{ab}$, and $\bold{\eta}_{Aab}$ is conjugate to $\ou{\psi}{A}{a}$. An obvious and important question is how the covariant phase space at the inner null surface relates to Ashtekar's phase space of radiative modes at null infinity \cite{AshtekarNullInfinity,Ashtekar:2014zsa}. This requires additional universal structures that only exist at $\mathcal{I}^\pm$, and we will come back to this question elsewhere. 

We then introduced the pre-symplectic two-form \eref{Omdef}, and studied its degenerate directions that define the gauge symmetries of the gravitational field in the bounded region $\mathcal{M}$ (see \hyperref[fig1]{figure 1}). We found that all internal $SL(2,\C)$ frame rotations are local gauge symmetries of the system, and the corresponding gauge generators vanish {on-shell}. At the boundary an additional internal gauge symmetry appeared: local $U(1)$ rotations of the boundary spinors are gauge transformations as well. Dilatations of the null normal sending $\ell^A$ and $\bold{\eta}_{Aab}$ into $\E^{-\lambda/2}\ell^A$ and $\E^{+\lambda/2}\bold{\eta}_{Aab}$  do, however, not define local gauge symmetries, they are genuine Hamiltonian motions, generated by the area Hamiltonian \eref{Arham}, whose on shell value is given by the area two-form smeared over a gauge parameter $\lambda:\mathcal{C}\rightarrow\R$. The most interesting case concerned, of course, the diffeomorphism symmetry, which is the genuine and indeed defining symmetry of general relativity. To sum up, we found the following: 
\begin{itemize}
\item[(i)] All four-dimensional diffeomorphisms $\varphi=\exp(\xi)$ that are generated by vector fields $\xi^a$ that vanish at the inner (resp.\ outer) boundary $\mathcal{N}$ (resp.\ $\mathcal{B}$) are gauge symmetries of the system.
\item[(ii)] Diffeomorphisms $\varphi=\exp(\xi)$ that are generated by vector fields $\xi^a|_{\mathcal{C}^{out}_{in}}$ $\in T\mathcal{C}^{in}_{out}$ that are tangential to the corners do not generate gauge transformations. The infinitesimal field variation $\mathcal{L}_\xi$ is not a degenerate direction of the pre-symplectic potential, and the interior product $\Omega_\varSigma(\mathcal{L}_\xi,\cdot)$ does not vanish \emph{on-shell}. Yet the infinitesimal field variation $\delta_\xi[\cdot]=\mathcal{L}_\xi(\cdot)$ is integrable and generated by a Hamiltonian, whose on-shell value is given by the diffeomorphism charges \eref{diffchargs}. If, in addition, $\xi^a_{(i)}=\uo{\epsilon}{ij}{k}x^k\partial^a_k+\mathcal{O}_{-}(\mathcal{\rho}^0)$ is an asymptotic rotation, its asymptotic charge \eref{ADMcharge} returns the ADM angular momentum at infinity. But there are infinitely many more charges \eref{surfchrge}, because there is an infinite functional freedom for $\xi^a\in T\mathcal{C}_{in}$. 
\item[(iii)] Finally, there are those diffeomorphisms $\exp(\xi)$ that drag\footnote{Diffeomorphisms that are generated by transversal vector fields $\xi^a|_{\mathcal{C}_{in}}\notin T\mathcal{N}$ fall out this framework. This is to be expected: By considering the gravitational field as a Hamiltonian system in a subregion $\mathcal{M}$, only those diffeomorphism can be Hamiltonian motions that preserve the subsystem and do not bring in new degrees of freedom. Hence $\xi^a\big|_{\mathcal{C}_{in}}\stackrel{!}{\in} T\mathcal{M}$} the partial Cauchy surface along the inner (outer) boundary (e.g.\ $\xi^a\big|_{\mathcal{C}_{in}}=\alpha\ell^a$). They do not generate gauge symmetries (i.e.\ $\Omega_\varSigma(\mathcal{L}_\xi,\cdot)\neq 0)$, and they are not integrable either: A quasi-local Hamiltonian $H_{\xi}$ exists if and only if the quasi-local first law $\delta H_{\xi}=\Omega_\varSigma(\delta,\mathcal{L}_\xi)$ is satisfied. If, in addition, we are considering a particular solution of Einstein's equations, for which $\xi^a$ is a Killing vector, the quasi-local first-law turns \eref{fstlaw} into the global first law \eref{fstlaw2} linking the ADM charges at infinity with the integral of the surface gravity over a cross-section of the inner null boundary.
\end{itemize}


\subsection{Relevance for quantum gravity}
Finally, let us discuss the relevance of the framework for quantum gravity. 
 The idea is simple: Given the symplectic structure \eref{CornerBrckt} at the inner corner, we introduce canonical creation and annihilation operators. This requires some additional background structures:  First of all, we pick a fiducial and non-degenerate two-dimensional volume element\footnote{$\ucheck{a},\ucheck{b},\dots$ are tensor indices at the inner corner.}
\begin{equation}
d^2\Omega\equiv{}^\circ\!\bold{\varepsilon}_{\ucheck{a}\ucheck{b}}
\end{equation}
at the corner, where ${}^\circ\!\bold{\varepsilon}_{\ucheck{a}\ucheck{b}}$ is the corresponding fiducial two-form. Notice, that $d^2\Omega$ can be seen either as a two-dimensional scalar density or a two-form, and this will become important in a moment. Let now ${}^\circ\!\bold{\varepsilon}^{\ucheck{a}\ucheck{b}}\in T\mathcal{C}\otimes T\mathcal{C}$ denote its inverse: ${}^\circ\!\bold{\varepsilon}^{\ucheck{a}\ucheck{c}}\,{}^\circ\!\bold{\varepsilon}_{\ucheck{b}\ucheck{c}}=\delta^{\ucheck{a}}_{\ucheck{b}}$, such that we can define the following \emph{momentum spinor}
\begin{equation}
\pi_A=\frac{\I}{16\pi G}{}^\circ\!\bold{\varepsilon}^{\ucheck{a}\ucheck{b}}\bold{\eta}_{A\ucheck{a}\ucheck{b}}.
\end{equation}
We may now also choose\footnote{The easiest and most geometrical way to choose such a normal is to use the surface normal $n^a|_{\mathcal{C}}\in T\mathcal{M}$ of the partial Cauchy surface $\varSigma$ itself, to say, in other words: $\delta_{AA'}|_{\mathcal{C}}=-\I\,\sqrt{2}\,e_{AA'a}n^a|_{\mathcal{C}}$.} an internal and future pointing four-vector $n^\alpha:n^\alpha n_\alpha=-1$ at the inner corner, such that the spin bundle over the corner is equipped with a Hermitian metric
\begin{equation}
\delta_{AA'}:=\sigma_{AA'\alpha}n^\alpha.
\end{equation}
 We can then define (in units of $\hbar=c=1$) the following spinor-valued Landau operators
\begin{subalign}
	a^A&=\frac{{d^{\frac{2}{2}}\Omega}}{\sqrt{2}}\left(\delta^{AA'}\bar{\ell}_{A'}-\I\pi^A\right),\\
b^A&=\frac{{d^{\frac{2}{2}}\Omega}}{\sqrt{2}}\left(\ell^A+\I\delta^{AA'}\bar\pi_{A'}\right),
\end{subalign}
where ${d^{\frac{2}{2}}\Omega}$ is the half-density 
 $d^{\frac{2}{2}}\Omega=\sqrt{d^2\Omega}$. The canonically conjugate operators are obtained by complex conjugation and lowering an index with the Hermitian metric, namely by
\begin{equation}
a^\dagger_A:=\delta_{AA'}\bar{a}^{A'},\qquad b^\dagger_A:=\delta_{AA'}\bar{b}^{A'}.
\end{equation}
The only non-vanishing canonical Poisson commutation relations are those for two pairs of harmonic oscillators, namely
\begin{equation}
\big\{a^A(z),a^\dagger_B(z')\big\}_{\mathcal{C}}=
\big\{b^A(z),b^\dagger_B(z')\big\}_{\mathcal{C}}=\I\delta^A_B\delta^{(2)}_{\mathcal{C}}(z,z').
\end{equation}
We can now build a Hilbertspace at the corner, simply by following the canonical procedure: The Fock vacuum at the sphere is the state in the kernel of the annihilation operators,
\begin{equation}
\widehat{a^A(z)}\big|0,\{d^2\Omega,n_\alpha\}\big\rangle=
\widehat{b^A(z)}\big|0,\{d^2\Omega,n_\alpha\}\big\rangle=0.\label{Fockvacuum}
\end{equation}
It is then easy to show that the reality conditions \eref{realcond} turn into the constraint that both oscillators have equal occupation numbers $N_a(z)=a^\dagger_A(z) a^A(z)$ and $N_b(z)=b^\dagger_A(z)b^A(z)$.

The construction of this vacuum state \eref{Fockvacuum} depends, however, crucially on an additional structure: It depends on a choice for an internal reference normal $n^\alpha$ and a fiducial area element $d^2\Omega$. What is then the operator that can measure or probe these labels, these say \qq{magnetic} quantum numbers $m=(d^2\Omega,n_\alpha)$?  A generic diffeomorphism $\exp(\xi):\mathcal{C}\rightarrow\mathcal{C}$ for a vector field $\xi^{a}\in T\mathcal{C}$ at the corner, will not preserve either of the fiducial structures, neither $d^2\Omega$ nor $n_\alpha$.  At the level of the classical phase space, such a diffeomorphism is generated by the Hamiltonian vector field of the surface charge $J_\xi[\mathcal{C}_{in}]$ (defined as in \eref{surfchrge} above), whose canonical quantisation will generate a representation of two-dimensional diffeomorphisms on the Fock space of the inner corner. In fact, we cannot expect that the Fock vacuum \eref{Fockvacuum} is invariant under such two-dimensional {diffeomorphisms}. This follows from our canonical analysis, for we saw (\hyperref[sec3]{section 3}) that the diffeomorphisms at the corner do not define degenerate directions of the pre-symplectic two-form $\Omega_\varSigma$, hence they are genuine Hamiltonian motions, and should not annihilate physical states. We expect, therefore, that the generalised angular moments send one Fock vacuum into the other, such that
\begin{equation}
\widehat{\exp\left(-\I J_\xi[\mathcal{C}]\right)}\big|0,\{d^2\Omega,n_\alpha\}\big\rangle\propto
\big|0,\{\exp(\xi)^\ast d^2\Omega,\exp(\xi)^\ast n_\alpha\}\big\rangle.
\end{equation}

An article is under preparation, where the resulting quantum theory is developed further and compared in particular with non-perturbative and covariant quantum gravity \cite{Ashtekar, thiemann, rovelli, status, alexreview}. So far, the results are promising: Upon introducing the Immirzi parameter, the quantisation of the total area of the inner corner reproduces the discrete loop quantum gravity area spectrum \cite{Rovelliarea, AshtekarLewandowskiArea}. The formalism seems to be relevant for other approaches to quantum gravity as well: The ambiguity of defining the Fock vacuum, which depends parametrically on a pair $\{n_\alpha,d^2\Omega\}$, is reminiscent of the IR ambiguities that appear in the definition of the radiative vacuum at null infinity \cite{AshtekarNullInfinity,Ashtekar:2014zsa}. These ambiguities play a crucial role for the definition of certain soft hairs that have been conjectured recently by Hawking, Perry and Strominger \cite{Hawking:2016msc} to carry the missing information in the process of Hawking evaporation. In addition, any choice for $\{n_\alpha,d^2\Omega\}$ is also related to certain gravitational edge modes that have been pointed out recently by Freidel, Donnelly and Perez \cite{Freidel:2015gpa,Donnelly:2016auv} for the gravitational field in bounded regions with inner corners.


\section*{Acknowledgments} 
I am most grateful to Abhay Ashtekar, Carlo Rovelli and Simone Speziale for the kind hospitality in Marseille and many engaging discussions during my visit to the Centre de Physique Theoriqué. 
This research was supported in part by Perimeter Institute for Theoretical Physics. 
Research at Perimeter Institute is supported by the Government of Canada through the Department of Innovation, Science and Economic Development and by the Province of Ontario through the Ministry of Research and Innovation.

\appendix
\section[Null geometry from boundary spinors]{Reconstruction of the intrinsic null geometry from the boundary spinors}\label{appdxA}
In the following, we show that the intrinsic geometry\footnote{Given by a pair $(q_{ab},[\ell^a])$ consisting of a degenerate signature $(0$$+$$+)$ metric $q_{ab}$ and an equivalence class $[\ell^a]$ of null directions.} $(q_{ab},[\ell^a])$ of the null surface $\mathcal{N}$  can be reconstructed from the boundary spinor $\bar{Z}_{\mathfrak{a}}=(\bold{\eta}_{Aab},\bar\ell^{A'})$ alone. First of all, we look at the inverse problem. Consider thus the pull back $(\varphi^\ast_{\mathcal{N}}e^{AA'})_a$ of the soldering form to a null surface $\mathcal{N}\hookrightarrow\mathcal{M}$. Adopting a Newman\,--\,Penrose notation, we write
\begin{equation}
(\varphi^\ast_{\mathcal{N}}e^{AA'})_a\equiv\ou{e}{AA'}{{\underleftarrow{a}}}=-\I \ell^A\bar{\ell}^{A'}k_a+\I\ell^A\bar{k}^{A'}\bar{m}_a+\I k^A\bar{\ell}^{A'}{m}_a,\label{epullb}
\end{equation}
where the spinors $k^A$ and $\ell^A$ are linearly independent and normalised to $k_A\ell^A=1$. The one-form 
$k_a\in T^\ast\mathcal{N}$ is real, and $m_a\in T^\ast_\C\mathcal{N}$ is complex, and together they form a basis $(k_a,m_a,\bar{m}_a)$ in the complexified co-tangent space $T^\ast_\C\mathcal{N}$. 
Equation \eref{epullb} clearly determines the pull-back of the Pleba\'nski two-form as well,
\begin{equation}
\Sigma_{AB\underleftarrow{ab}}=-e_{AC'\underleftarrow{a}}e_{B}{}^{C'}{}_{\underleftarrow{b}}=2\ell_A\ell_B k_{[a}\bar{m}_{b]}+2\ell_{(A}k_{B)}\bar{m}_{[a}{m}_{b]}.
\end{equation}
Going back to the glueing condition \eref{glu}, we can now identify the spinor-valued two-form $\bold{\eta}_{Aab}$ on $\mathcal{N}$ as 
\begin{equation}
\bold{\eta}_{Aab}=2\ell_Ak_{[a}\bar{m}_{b]}+2k_A\bar{m}_{[a}m_{b]}.\label{mudecomp0}
\end{equation}
Notice that there is a gauge ambiguity in the decomposition: Given the pull-back of the tetrad $\ou{e}{AA'}{\underleftarrow{a}}$ on $\mathcal{N}$, 
the boundary spinor $(\bold{\eta}_{Aab},\bar\ell^{A'})$ can be determined only up to dilatations (e.g.\ $\ell^A\rightarrow\E^{\lambda}\ell^A$) and $U(1)$ phase transformations, which act collectively as
\begin{equation}
(\bold{\bar\eta}_{A'ab},\ell^A)\longrightarrow(\E^{-\frac{\lambda-\I\phi}{2}}\bold{\bar\eta}_{A'ab},\E^{+\frac{\lambda+\I\phi}{2}}\ell^A).
\end{equation}
for local gauge parameters $\lambda,\varphi\in\R$.\vspace{1em} 

Next, we look at the inverse problem: Given a surface spinor $(\bold{\eta}_{Aab},\bar\ell^{A'})$, which satisfies the reality condition \eref{realcond}, we want to recover the intrinsic geometry of the null surface $\mathcal{N}$. The construction is as follows: First of all, we extend $\ell^A$ with a second linearly independent spinor $k^A$ into a normalised dyad $(k^A,\ell^A):k_A\ell^A=1$ and decompose $\bold{\eta}_{Aab}$ into complex-valued two-forms $\bold{\mu}_{ab}$ and $\bold{\varepsilon}_{ab}$ on the three-manifold $\mathcal{N}$, namely
\begin{equation}
\bold{\eta}_{Aab}=\bold{\mu}_{ab}\ell_A+\I\bold{\varepsilon}_{ab}k_A.\label{mudecomp1}
\end{equation}
The reality condition \eref{realcond} implies that the two-form $\bold{\varepsilon}_{ab}$ is real. Now, $\bold{\varepsilon}_{ab}$ is a two-form in three dimensions, which implies that there is at least one degenerate direction. In fact, there can be only one, because otherwise the intrinsic geometry of $\mathcal{N}$ is completely degenerate. This one unique degenerate direction determines an equivalence class $[\ell^a]$ of null generators on $\mathcal{N}$. In other words
\begin{equation}
T\mathcal{N}\ni l^a\in[\ell^a]\Leftrightarrow \bold{\eta}_{Aab}\ell^A l^a=0.\label{eqclass}
\end{equation}
The null surface is orientable, hence there is the \emph{metric-independent} Levi-Civita density $\oepsilon^{abc}$ (of density weight one), and the inverse density $\uepsilon_{abc}$, that we may use to define the following vector-valued densities on $\mathcal{N}$, namely
\begin{equation}
\bold{\ell}^a=\frac{1}{2}\oepsilon^{abc}\bold{\varepsilon}_{bc},
\end{equation}
which is real, and 
\begin{equation}
\bold{\bar\mu}^a=\frac{\I}{2}\oepsilon^{abc}\bold{\mu}_{bc},
\end{equation}
	which is complex. A short moment of reflection reveals that $\bold{\ell}^a$ lies itself in the equivalence class \eref{eqclass}. We now have a triple
\begin{equation}
(\bold{\ell}^{a},\bold{\mu}^a,\bold{\bar\mu}^a)\label{nutriad}
\end{equation}
of density-valued tangent vectors on $\mathcal{N}$. In general, this triple of vectors will be linearly independent (if it were linearly dependent, the intrinsic geometry of $\mathcal{N}$ would be degenerate, and there would be no non-degenerate spacetime metric in a neighbourhood of the three-boundary $\mathcal{N}$). Assume, in addition, that the triad \eref{nutriad} is positively oriented\footnote{We mean that the density $\I\uepsilon_{abc}\bold{\ell}^a\bold{\mu}^b\bold{\bar\mu}^c$ is positive. The case of opposite orientation can be studied in complete analogy.}. Next, define a dual basis $(k_a,m_a,\bar{m}_a)$ of $T^\ast_\C\mathcal{N}$ by declaring that the only non-vanishing interior products between $(\bold{\ell}^{a},\bold{\mu}^a,\bold{\bar\mu}^a)$ and $(k_a,m_a,\bar{m}_a)$ simply be
\begin{equation}
\bold{\mu}^a\bar{m}_a=\bold{\bar\mu}^a{m}_a=-\bold{\ell}^ak_a=\bold{N},
\end{equation}
for some positive normalisation $\bold{N}$, which is a density of weight one. The two bases are not independent, the densitised triad \eref{nutriad}  can be obtained always from $(k_a,m_b,\bar{m}_c)$ by dualisation, namely by
\begin{subalign}
\bold{\ell}^a&=-\bold{N}\frac{\oepsilon^{abc}m_b\bar{m}_c}{\oepsilon^{def}k_dm_e\bar{m}_f},\\
\bold{\mu}^a&=+\bold{N}\frac{\oepsilon^{abc}k_b{m}_c}{\oepsilon^{def}k_dm_e\bar{m}_f}.
\end{subalign}
The normalisation can be finally fixed by the requirement
\begin{equation}
\bold{\varepsilon}_{ab}\stackrel{!}{=}2\I m_{[a}\bar{m}_{b]},
\end{equation}
which implies
\begin{equation}
\bold{\mu}_{ab}=2k_{[a}\bar{m}_{b]}.
\end{equation}
Comparing the last two equations with the decomposition \eref{mudecomp0} of $\bold{\eta}_{Aab}$ into component functions $\bold{\mu}_{ab}$ and $\bold{\varepsilon}_{ab}$, we thus see that a generic configuration of $\bar{Z}_{\mathfrak{a}}=(\bold{\eta}_{Aab},\b\ell^{A'})$, which satisfies the reality condition \eref{realcond}, will endow the three-manifold $\mathcal{N}$ with a null direction $[\ell^a]$ and a degenerate signature $(0$$+$$+)$ metric
\begin{equation}
q_{ab}=2m_{(a}\bar{m}_{b)}.
\end{equation}
Our fundamental configuration variable on the three-manifold $\mathcal{N}$ is, therefore, the boundary spinor $\bar{Z}_{\mathfrak{a}}=(\bold{\eta}_{Aab},\bar\ell^{A'})$. The area two-form $\bold{\varepsilon}_{ab}$, the equivalence class $[\ell^a]$ of null generators and the signature $(0$$+$$+)$ metric $q_{ab}$ are secondary, for they can be all reconstructed from $\bold{\eta}_{Aab}$ and $\ell^A$ alone.
\section[ADM energy from corner term at infinity]{ADM energy from the corner term at infinity}
For the completeness of the presentation, I will briefly sketch (using the methods of this paper) the very well know result that asymptotic time translations are Hamiltonian, that, in other words, the functional variation of the ADM energy is obtained from the functional interior product of the pre-symplectic two-form \eref{Omdef} with the field variation $\delta_\xi=\mathcal{L}_\xi$ for an asymptotic time translation\footnote{For a more complete analysis, see \cite{Corichi:2015cqa,Ashtekar:2008jw} and references in there. } $\xi^a=t^a$. 
The starting point is equation \eref{Hvar}, which now becomes
\begin{align}\nonumber
\Omega_\varSigma&(\delta,\mathcal{L}_\xi)=(\text{terms at the inner corner})+\\\nonumber
&-\frac{1}{8\pi G}\int_{\mathcal{C}_{out}}\!\!\Big(\delta(\ast\Sigma_{\alpha\beta}) z^\alpha\mathcal{L}_\xi z^\beta-\mathcal{L}_\xi(\ast\Sigma_{\alpha\beta})\wedge z^\alpha\delta z^\beta+\\
&\hspace{14em}+\frac{1}{2}(\xi\hook\ast\!\Sigma_{\alpha\beta})\wedge {\delta A}^{\alpha\beta}\Big).\label{Hvar2}
\end{align}%
We will evaluate this integral in the limit, where the outer corner is sent to spacelike infinity. Let us specify this limit more carefully. Consider thus asymptotically inertial coordinates $\{x^\alpha\}=\{t,x^i\}$. Define the hyperbolic distance $\rho=\sqrt{\eta_{\alpha\beta}x^\alpha x^\beta}$ (for $|t|<|\vec{x}|$) and define (for every $\rho$) the outer and timelike cylinder $\mathcal{B}_\rho$ as an $\rho=\mathrm{const}.$ hypersurface. Let $z^\alpha=\ou{e}{\alpha}{a}z^a$ be the internal and outwardly oriented normal to this cylinder, such that 
\begin{equation}
\varphi^\ast_{\mathcal{C}_{\rho}}(e_\alpha z^\alpha)=0,
\end{equation}
where $\varphi_{\mathcal{C}_{\rho}}^\ast:T^\ast\mathcal{M}\rightarrow T^\ast\mathcal{C}_{\rho}$ denotes the pull-back to the outer corner, which is a two-dimensional cross-section of the boundary $\mathcal{B}_\rho$. For $\rho\rightarrow\infty$, there exists then the following asymptotic expansion of the tetrad
\begin{equation}
\ou{e}{\alpha}{a}=
\nabla_a(\rho z^\alpha)+\ou{f}{\alpha}{a},\label{expnsn}
\end{equation}%
where $\nabla_a z^\alpha=\partial_a z^\alpha+\ou{A}{\alpha}{\beta a}z^{\beta}$ is the covariant derivative with respect to the spin connection, and the perturbation $\ou{f}{\alpha}{a}$ is $\mathcal{O}(\rho^{-1})$ even: the limit $\ou{f}{\mu}{\nu}(\eta,\vartheta,\varphi):=\lim_{\rho\rightarrow\infty} \rho (\ou{f}{\mu}{a}\partial^a_\nu)(\eta,\rho,\vartheta,\varphi)$ exist and defines an even function on the two-sphere at infinity ($\vartheta$, $\varphi$ are spherical coordinates, $\eta$ is an hyperbolic angle). 

The first two terms in the second line of \eref{Hvar2} vanish due to the parity and falloff conditions for $\delta\ou{e}{\alpha}{a}=\delta \ou{f}{\alpha}{a}$, which is $\mathcal{O}(\rho^{-1})$ and parity even. We are thus left with the last term, and we evaluate it using the asymptotic expansion \eref{expnsn} of the tetrad. The only term that has a non-vanishing limit for $\rho\rightarrow\infty$ is
\begin{align}
\lim_{\rho\rightarrow\infty}\int_{\mathcal{C}_\rho}\epsilon_{\alpha\beta\mu\nu}\xi^\mu e^\nu\wedge\delta A^{\alpha\beta}=
\lim_{\rho\rightarrow\infty}\int_{\mathcal{C}_\rho}\epsilon_{\alpha\beta\mu\nu}\xi^\mu \nabla(\rho z^\nu)\wedge\delta A^{\alpha\beta},
\end{align}
where $\xi^\alpha=\ou{e}{\alpha}{a}\xi^a$. Next, we use Stokes's theorem to shuffle the exterior covariant derivative from $z^\mu$ to $\delta A^{\alpha\beta}$. The term where the exterior derivative hits $\xi^\alpha$ vanishes, because $\xi^a$ is an asymptotic Poincaré translation and its derivative $\uo{e}{\beta}{b}\nabla_b\xi^\alpha$ vanishes for $\rho\rightarrow\infty$ as $\rho^{-2}$ or faster.  Now, we also know that the variation of the field strength $\delta F^{\alpha\beta}$ is given by the exterior covariant derivative of the variation of the connection: $\delta F^{\alpha\beta}=\nabla\delta A^{\alpha\beta}$, hence
\begin{equation}
\lim_{\rho\rightarrow\infty}\int_{\mathcal{C}_\rho}\epsilon_{\alpha\beta\mu\nu}\xi^\mu e^\nu\wedge\delta A^{\alpha\beta}=-\lim_{\rho\rightarrow\infty}\int_{\mathcal{C}_\rho}\rho\,\epsilon_{\alpha\beta\mu\nu}\xi^\mu z^\nu \delta F^{\alpha\beta}.\label{Energy1}
\end{equation}
Given the vacuum Einstein equations, the field strength of the $SO(1,3)$ spin connection can be written in terms of the Weyl tensor,
\begin{equation}
\ou{F}{\alpha}{\beta cd}=e^{\alpha a}\uo{e}{\beta}{b}C_{abcd}.
\end{equation}
In the limit of $\rho\rightarrow\infty$, only the variation of the electric component
\begin{equation}
E_{ab}=C_{cadb}z^cz^d\label{Epart}
\end{equation}
 of the Weyl tensor $C_{abcd}$ contributes to the integral \eref{Energy1} of $\delta F^{\alpha\beta}$ over the two-surface at infinity, and we can rewrite, therefore, the integral \eref{Energy1} in terms of $E_{ab}$ alone,
 \begin{equation}
\lim_{\rho\rightarrow\infty}\int_{\mathcal{C}_\rho}\epsilon_{\alpha\beta\mu\nu}\xi^\mu e^\nu\wedge\delta A^{\alpha\beta}=2\lim_{\rho\rightarrow\infty}\int_{\mathcal{C}_\rho}d^2\Omega\,\rho^3n^a\xi^b\delta E_{ab},\label{ADMcharge0}
\end{equation}
where we introduced the fiducial area element $d^2\Omega=\rho^{-2}\ast\!\Sigma_{\alpha\beta}n^\alpha z^\beta$ on the two-sphere at infinity, with $n^\alpha:n_\alpha z^\alpha=0$ denoting the future oriented and internal normal to both $z^\alpha$ and $\mathcal{C}_\rho$ ($\varphi^\ast_{\mathcal{C}_\rho}e_\alpha n^\alpha=0$). In the Ashtekar\,--\,Hansen framework \cite{Ashtekar:1978zz} of asymptotic infinity, the charge $P_\alpha\xi^\alpha$ for an asymptotic translation $\xi^a$ is defined as the integral
\begin{equation}
P_\alpha\xi^\alpha=\frac{1}{8\pi G}\lim_{\rho\rightarrow\infty}\int_{\mathcal{C}_\rho}d^2\Omega\,\rho^3n^a\xi^b E_{ab},\label{ADMcharge1}
\end{equation}
of the $\mathcal{O}(\rho^{-3})$ even electric part \eref{Epart} of the Weyl tensor over a two-sphere at infinity. The only functional dependence of this asymptotic charge is in $E_{ab}$. The two-dimensional volume element $d^2\Omega$, the radial and hyperbolic coordinate $\rho\rightarrow\infty$ and the time-like normal vector $n^a=\uo{e}{\alpha}{a}n^\alpha$ are fiducial background structures.  
Equation \eref{ADMcharge0} can be expressed, therefore, as a total variation of the charge \eref{ADMcharge1} at spacelike infinity. This charge is nothing but the ADM four-momentum contracted with $\xi^\alpha$, and we have thus shown that the Hamiltonian vector field of the canonical ADM charges \eref{ADMcharge} generate asymptotic time translations, which is the same as to say
\begin{align}
\Omega_\varSigma&(\delta,\mathcal{L}_\xi)=(\text{terms at the inner corner})-\xi^\alpha\delta P_\alpha.\label{Hvar3}
\end{align}%


\providecommand{\href}[2]{#2}\begingroup\raggedright\endgroup

\end{document}